\documentclass[12pt,english]{article}
\usepackage{lmodern}
\usepackage[T1]{fontenc}
\usepackage[latin9]{inputenc}
\usepackage{amsmath}
\usepackage{amssymb}
\usepackage{esint}

\usepackage{geometry}
\makeatletter
\usepackage{float} % allows forcing plot and table locations with [H]
\usepackage{hyperref}
\usepackage{cite}
\usepackage{babel}
\usepackage{mathdots}
\usepackage{amsfonts}
\usepackage{mathrsfs}
\usepackage{amsmath}
\usepackage{esint}
\usepackage{graphicx}
\usepackage{youngtab}
\usepackage{multicol}
\usepackage{multirow}
\usepackage{slashed}
\usepackage{bm}
\usepackage{relsize}
\usepackage[dvipsnames]{xcolor}

\allowdisplaybreaks

\makeatother

\usepackage{babel}

\begin{document}

\author{Kai-Yu Zhang\footnote{Email: kaiyu\_zhang@tju.edu.cn}~~and Wan-Zhe Feng\footnote{Email: vicf@tju.edu.cn, corresponding author}\\
\textit{\small Center for Joint Quantum Studies and Department of Physics,}\\
\textit{\small School of Science, Tianjin University, Tianjin 300350, PR. China}}

\title{\LARGE  Explaining the $W$ boson mass anomaly and dark matter with a $U(1)$ dark sector}

\date{}
\maketitle

\begin{abstract}

  The $W$ boson mass recently reported by the CDF collaboration
  shows a deviation from the standard model prediction
  with an excess at $7\sigma$ level.
  We investigate two simple extensions of the standard model with an extra $U(1)$ dark sector.
  One is the $U(1)_x$ extension, where the $U(1)_x$ gauge field mixes with the standard model through gauge kinetic terms.
  The other is a general $U(1)_{\mathbf{A} Y+\mathbf{B} q}$ extension of the standard model.
  Fitting various experimental constraints we find the $U(1)_x$ extension with only kinetic mixing
  can enhance the $W$ boson mass for at most 10~MeV.
  While the $U(1)_{\mathbf{A} Y+\mathbf{B} q}$ extension can easily generate 77~MeV enhancement of the $W$ boson mass
  and also offer a viable dark matter candidate with mass ranging from several hundred GeV to TeV,
  which may be detected by future dark matter direct detection experiments with improved sensitivities.

\end{abstract}

\newpage

\tableofcontents

\section{Introduction}\label{sec:Intro}

The CDF collaboration recently reported a direct measurement of the $W$ boson mass with increased precision~\cite{CDF:2022hxs}
\begin{equation}
M_W^{\rm CDF} =  80.4335 \pm 0.0094~{\rm GeV}\,,
\end{equation}
which has a deviation from the Standard Model (SM) expectation~\cite{ParticleDataGroup:2020ssz}
\begin{equation}
M_W^{\rm SM} =  80.357 \pm 0.006~{\rm GeV}
\end{equation}
at a confidence level of $7\sigma$.
This result soon attracts a lot of discussions and explorations in particle physics~\cite{Fan:2022dck,Zhu:2022tpr,Lu:2022bgw,Athron:2022qpo,Yuan:2022cpw,Strumia:2022qkt,
Yang:2022gvz,deBlas:2022hdk,Du:2022pbp,Tang:2022pxh,Cacciapaglia:2022xih,Blennow:2022yfm,
Arias-Aragon:2022ats,Zhu:2022scj,Sakurai:2022hwh,Fan:2022yly,Liu:2022jdq,Lee:2022nqz,Cheng:2022jyi,
Song:2022xts,Bagnaschi:2022whn,Paul:2022dds,Bahl:2022xzi,Asadi:2022xiy,DiLuzio:2022xns,Athron:2022isz,
Gu:2022htv,Heckman:2022the,Babu:2022pdn,Heo:2022dey,Du:2022brr,Cheung:2022zsb,DiLuzio:2022ziu,
Crivellin:2022fdf,Endo:2022kiw,Biekotter:2022abc,Balkin:2022glu,Krasnikov:2022xsi,Ahn:2022xeq,Han:2022juu,Zheng:2022irz,
Kawamura:2022uft,Peli:2022ybi,Ghoshal:2022vzo,Perez:2022uil,Kanemura:2022ahw}.
The recent CDF result is however in tension with previous measurements on the $W$ boson mass
from other experimental groups~\cite{D0:2012kms,ALEPH:2013dgf,ATLAS:2017rzl,LHCb:2021bjt},
and needs to be further checked with future LHC measurements.
At the moment, details of the CDF measurements such as calibrations and experimental uncertainties,
as well as details of data analysis like selection rules and fitting assumptions that CDF is using,
need to be better understood before one can make any conclusive statement on the CDF new result.
Nevertheless, this intriguing result still points to new physics beyond the SM.
In this paper, we discuss a possible explanation of the $W$ boson mass anomaly
as well as the nature of dark matter with an extra $U(1)$ dark sector.

Dark sectors with new interactions and new hypothetical particles
are usually introduced to explain puzzles beyond the SM.
Among them,
$U(1)$ dark sectors are the simplest and are well-motivated from grand unified theories and string theory~\cite{Feng:2014cla,Feng:2014eja,Anchordoqui:2014wha}.
The dark $U(1)$ gauge field may mix with the $U(1)$ hypercharge via gauge kinetic terms~\cite{Holdom:1985ag,Feldman:2007wj,Aboubrahim:2020lnr}.
However we find for this case,
such kinetic mixing can only generate at most 10~MeV mass enhancement to the $W$ boson
according to various experimental constraints.
While the $U(1)_{\mathbf{A} Y+\mathbf{B} q}$ extension of the SM
can easily generate 77~MeV enhancement of the $W$ boson mass, explaining the CDF new result.
In addition, fermions only charged under the extra $U(1)$ gauge group are natural dark matter candidates.
The massive neutral vector bosons of the theory will act as vector portal between the dark sector and SM particles.
The dark fermion can in principle annihilate through vector bosons exchange into SM fermion pairs
and satisfy the current observed value of the dark matter relic density.

The paper is organized as follows.
In Section.~\ref{sec:darkU1} we introduce the $U(1)_x$ extension of the SM,
and explain the kinetic mixing between the $U(1)_x$ dark sector and the SM.
In Section.~\ref{sec:generalU1} we discuss the $U(1)_{\mathbf{A} Y+\mathbf{B} q}$ extension of the SM.
In Section.~\ref{sec:WM} we review the $S,T,U$ effective Lagrangian approach
and calculate the effective shifts in the oblique parameters for the two $U(1)$ extensions of the SM,
which are essential in explaining the $W$ boson mass enhancement.
Fitting various experimental constraints, we investigate
how these two models explain the $W$ boson mass enhancement.
We then focus on the fermionic $U(1)$ dark matter candidate
in the $U(1)_{\mathbf{A} Y+\mathbf{B} q}$ extension of the SM in Section.~\ref{sec:DM}.
We calculate the dark matter relic abundance and fit our benchmark points with direct and indirect detection bounds.
Finally we conclude in Section.~\ref{sec:Con}.

\section{The $U(1)_x$ extension of the SM with kinetic mixing}\label{sec:darkU1}

We first briefly discuss the dark $U(1)_x$ extension of the SM.
The mixing of the $U(1)$ dark sector with the SM can be generated
through either the gauge kinetic terms~\cite{Holdom:1985ag} or the mass terms~\cite{Kors:2004dx,Kors:2005uz,Feldman:2006wb}.
In this section we will only discuss the kinetic mixing effect~\cite{Holdom:1985ag,Aboubrahim:2020lnr}.
The kinetic terms of the gauge fields are given by
\begin{equation}
\mathcal{L}=-\frac{1}{4}W^a_{\mu\nu}W^{a\,\mu\nu}
-\frac{1}{4}B_{\mu\nu}B^{\mu\nu}
-\frac{1}{4}C_{\mu\nu}C^{\mu\nu}-\frac{\delta}{2}B_{\mu\nu}C^{\mu\nu}\,,
\end{equation}
where $W^a_{\mu \nu},B_{\mu \nu},C_{\mu \nu}$ are the $SU(2)_L,U(1)_Y,U(1)_x$ field strengths,
and $\delta$ is the kinetic mixing parameter.
Since we only consider the kinetic mixing effect in this section,
we will not go into details of the $U(1)_x$ breaking mechanism.
The $U(1)_x$ gauge field $C_\mu$ can obtain a mass $M_1$ from
either Higgs mechanism, or Stueckelberg mechanism.
In the gauge eigenbasis $V^{T}=(C,B,A^{3})$,
the $U(1)_x$ gauge boson $C$ mixes with the SM gauge bosons via the following matrices
\begin{equation}
\mathcal{K}=\left(\begin{array}{ccc}
1 & \delta & 0\\
\delta & 1 & 0\\
0 & 0 & 1
\end{array}\right)\,,\qquad M^{2}=\left(\begin{array}{ccc}
M_{1}^{2} & 0 & 0\\
0 & \frac{1}{4}v^{2}g_{Y}^{2} & -\frac{1}{4}v^{2}g_{2}g_{Y}\\
0 & -\frac{1}{4}v^{2}g_{2}g_{Y} & \frac{1}{4}v^{2}g_{2}^{2}
\end{array}\right)\,.
\end{equation}
A simultaneous diagonalization of both the kinetic mixing matrix $\mathcal{K}$
and the mass-squared matrix $M^{2}$ leads to the relation between
the mass eigenbasis $E^{T}=(Z^{\prime},A_{\gamma},Z)$ and the original gauge eigenbasis as $V=RE$,
where $R$ is the transformation matrix, given by
\begin{align}
	\left(\begin{array}{c}
		C\\
		B\\
    	A^{3}
	\end{array}\right) & =
	\left(\begin{array}{ccc}
				c_{\delta}\cos\psi & 0 & c_{\delta}\sin\psi\\
				-s_{\delta} \cos\psi + s_W \sin\psi & c_W &  -\cos\psi s_W -s_{\delta} \sin\psi\\
				- c_W \sin\psi  & s_W & c_W \cos\psi
	\end{array}\right)\left(\begin{array}{c}
				Z^{\prime}\\
				A_\gamma\\
	        	Z
			\end{array}\right)\,, \label{Trans33}
\end{align}
where $c_{\delta}=1/\sqrt{1-\delta^{2}}$ and $s_{\delta}=\delta/\sqrt{1-\delta^{2}}$,
$s_W=\sin \theta_W$, $c_W=\cos \theta_W$, and the mixing angle $\psi$ is given by
\begin{equation}
	\tan2\psi=\frac{2\delta\sqrt{1-\delta^{2}}\sin\theta_{W}}{1-\delta^{2}(1+\sin^{2}\theta_{W}) -M_{1}^2/M_{0}^2}
\approx - 2 \delta \epsilon^2 \sqrt{1-\delta^{2}} s_W\,,\label{33Psi}
\end{equation}
where $M_0 = \frac{v}{2} \sqrt{g_2^2 + g_Y^2}$ is the $Z$ boson mass in the SM.
In this paper we will focus on the $Z^\prime$ mass region of the order of TeV scale and
we further define a parameter $\epsilon^2 \equiv M_0^2/M_1^2 \ll 1$.
The above diagonalization leads to a massless photon, and massive $Z$, $Z^\prime$ gauge bosons with masses
\begin{align}
M_Z^2 &= M_0^2 \,(1 - \delta^2 \epsilon^2 s_W^2 + \cdots )\,, \label{ZMco} \\
M_{Z^\prime}^2 &= M_1^2 \, (1 + \delta^2 +\cdots ) \,.
\end{align}

The $U(1)_x$ dark sector in general includes dark fermions.
The simplest case is to consider a Dirac fermion $\chi$ charged under $U(1)_x$ but not charged under the SM gauge groups.
The dark fermion part of the Lagrangian can be written as
\begin{equation}
\mathcal{L}_\chi= \bar{\chi} (i\gamma^\mu \partial_\mu - m_\chi)\chi - g_x Q_\chi \bar{\chi} \gamma^\mu \chi C_\mu\,,\label{DF}
\end{equation}
where the dark fermion $\chi$ has mass $m_\chi$ and carries the $U(1)_x$ charge $Q_\chi$.

The interactions of gauge bosons and fermions can be obtained using the transformation matrix $R$
\begin{equation}
-\mathcal{L}_{{\rm int}}=\left(g_{x}J_{x},g_{Y}J_{Y},g_{2}J_{3}\right)V=\left(g_{x}J_{x},g_{Y}J_{Y},g_{2}J_{3}\right)RE\,.
\end{equation}
In the mass eigenbasis $Z$ and $Z^\prime$ gauge bosons couple to dark fermions and all SM fermions,
while the photon has exactly zero coupling with dark fermions.
After the mixing, $Z$ boson to SM fermion couplings are modified to be
\begin{align}
	\mathcal{L}_{Z\bar{f}f}&=-\big(R_{23}g_{Y}J_{Y}^{\mu}+R_{33}g_{2}J_{3}^{\mu}\big)Z_\mu \\
	&\approx -\frac{e}{2s_W c_W} \bar{f}_i\gamma^\mu \left\{
	\big[
	( 1 -  \delta^2 \epsilon^2 s_W^2 )
	T_3^i -2Q^i s_W^2 (1- \delta^2 \epsilon^2 )
	\big]- ( 1 - \delta^2 \epsilon^2 s_W^2) T_3^i \gamma^5
	\right\} f_i Z_\mu\,, \nonumber
\end{align}
where $Q^i, T_3^i$ are respectively the electric charge and the third component of weak isospin
of the SM fermions.
The $Z$ boson also couples to dark sector fermions
\begin{equation}
   \mathcal{L}_{Z\bar{\chi}\chi}=-R_{13}g_{x}J_{x}^{\mu} Z_\mu
   \approx -\delta \epsilon^2 s_W g_x Q_x  \bar{\chi} \gamma^\mu \chi  Z_\mu \,.\label{ZHFcoup}
\end{equation}
The couplings of $Z^\prime$ gauge boson to fermions are worked out as
\begin{align}
   \mathcal{L}_{Z^{\prime}\bar{f}f}&=-\big(R_{11}g_{x}J_{x}^{\mu}+
   R_{21}g_{Y}J_{Y}^{\mu}+R_{31}g_{2}J_{3}^{\mu}\big)Z_\mu^{\prime}\\ \nonumber
   & \approx - g_x Q_x \bar{\chi} \gamma^\mu \chi Z_\mu^{\prime}
   + \frac{1}{2} \bar{f}_i \big[
   \delta g_Y (2Q^i - {T^i_3} )\gamma^\mu + \delta g_Y {T^i_3} \gamma^\mu \gamma^5\big] f_i
   Z_\mu^{\prime}\,.\label{ZPSMcoupK}
\end{align}
The couplings of neutral gauge bosons $Z,Z^\prime$ to SM fermions can be written in the standard form as
\begin{align}
    \mathcal{L}_{Z Z^\prime}=-\frac{e}{2s_W c_W}
    \big[\bar{f}_i\gamma^\mu (v_i-a_i\gamma^5 )f_i Z_\mu
    +\bar{f}_i\gamma^\mu (v_i^\prime-a_i^\prime\gamma^5)f_i Z^\prime_\mu
    \big]
\end{align}
where
\begin{align}
v_i &= T_3^i(\cos \psi + s_\delta s_W \sin \psi) -2Q_i s_W^{2}(\cos \psi+s_\delta s_W^{-1} \sin \psi) \,,\\
a_i &= T_3^i(\cos \psi + s_\delta s_W \sin \psi) \,, \\
v_i^{\prime} &=- T_3^i(\sin \psi-  s_\delta s_W \cos \psi) + 2Q_i s_W^{2}(\sin \psi- s_\delta s_W^{-1}\cos \psi) \,,\label{M1Zpvcou}\\
a_i^{\prime} &=- T_3^i(\sin \psi-  s_\delta s_W \cos \psi) \,.\label{M1Zpacou}
\end{align}
We notice that in the expression of $v_i,a_i$ the change to the $Z\bar f f$ coupling is proportional to $\delta^2 \epsilon^2 \lesssim 10^{-6}$ which is just a tiny modification.

\section{The $U(1)_{\mathbf{A} Y+\mathbf{B} q}$ extension of the SM}\label{sec:generalU1}

We now discuss a general $U(1)_{\mathbf{A} Y+\mathbf{B} q}$ extension of the SM,
where the $U(1)$ charge $Q=\mathbf{A} Y+\mathbf{B} q$ is
a linear combination of the hypercharge $Y$ and any possible charge
$q$, and $\mathbf{A},\mathbf{B}$ are free parameters. For example, the combination
$\mathbf{A} Y+\mathbf{B}(B-L)$ is the most general anomaly-free combination
with the inclusion of three generations of right-handed neutrinos.
The choice of $q$ can be family-dependent,
e.g., $L_{\mu}-L_{\tau}$~\cite{Feng:2012jn},
$B_{1}+B_{2}-2B_{3}$~\cite{Celis:2016ayl},
and can also be any hidden quantum number that SM fields are not charged.
This $U(1)_{\mathbf{A} Y+\mathbf{B} q}$ symmetry is broken by a complex
scalar field $\phi$ at some scale higher than the electroweak scale.
The complex scalar $\phi$ does not carry any SM gauge charges.
The covariant derivatives acting on the SM Higgs doublet $\Phi$ and the complex scalar $\phi$
are written as
\begin{align}
D_{\mu}\Phi & =\Big(\partial_{\mu}-ig_{2}T^{a}A_{\mu}^{a}-i\frac{1}{2}g_{Y}YB_{\mu}
-i\frac{1}{2}g_{Y}\mathbf{A} YC_{\mu}\Big)\,\Phi\,,\\
D_{\mu}\phi & =(\partial_{\mu}-ig_{x}QC_{\mu})\,\phi\,,
\end{align}
where $A_{\mu}^{a},B_{\mu},C_{\mu}$ are $SU(2)_L,U(1)_{Y},U(1)_{\mathbf{A} Y+\mathbf{B} q}$
gauge fields respectively.
The vevs of the scalar fields are given by
\begin{equation}
\langle\Phi\rangle=\left(\begin{array}{c}
0\\
\frac{v}{\sqrt{2}}
\end{array}\right)\,,\qquad\qquad\langle\phi\rangle=\frac{u}{\sqrt{2}}\,.
\end{equation}
The scalar fields $\Phi,\phi$ can mix via the scalar potential of the following
form
\begin{equation}
V(\Phi,\phi)=m_{\Phi}^{2}\Phi^{\dagger}\Phi+\mu^{2}|\phi|^{2}+\lambda_{1}(\Phi^{\dagger}\Phi)^{2}+\lambda_{2}|\phi|^{4}+\lambda_{3}\Phi^{\dagger}\Phi|\phi|^{2}\,,
\end{equation}
and we assume the scalar mixing is small due to the choice of the
parameters.

We are interested in the SM Higgs does not carry the $q$ quantum number.
In this case the neutral gauge bosons are mixed through the mass matrix
\begin{equation}
M^{2}=\left(\begin{array}{ccc}
g_{x}^{2}u^{2}+\frac{1}{4}v^{2}g_{\mathbf{A}}^{2} & \frac{1}{4}v^{2}g_{\mathbf{A}}g_{Y} & -\frac{1}{4}v^{2}g_{\mathbf{A}}g_{2}\\
\frac{1}{4}v^{2}g_{\mathbf{A}}g_{Y} & \frac{1}{4}v^{2}g_{Y}^{2} & -\frac{1}{4}v^{2}g_{2}g_{Y}\\
-\frac{1}{4}v^{2}g_{\mathbf{A}}g_{2} & -\frac{1}{4}v^{2}g_{2}g_{Y} & \frac{1}{4}v^{2}g_{2}^{2}
\end{array}\right)\,,
\end{equation}
where we define $g_{\mathbf{A}}\equiv \mathbf{A} g_{Y}$. A diagnolization
of the above mass matrix gives rise to a change of the gauge boson
basis into the mass eigenbasis as
\begin{equation}
\left(\begin{array}{c}
C\\
B\\
A^{3}
\end{array}\right)=\left(\begin{array}{ccc}
\cos\psi & 0 & -\sin\psi\\
-s_{W}\sin\psi & c_{W} & -s_{W}\cos\psi\\
c_{W}\sin\psi & s_{W} & c_{W}\cos\psi
\end{array}\right)\left(\begin{array}{c}
Z^{\prime}\\
A_{\gamma}\\
Z
\end{array}\right)\,.\label{Trans33G}
\end{equation}
The angle $\psi$ is given by
\begin{equation}
\tan2\psi=\frac{4g_{\mathbf{A}}vM_{0}}{4M_{0}^{2}-g_{\mathbf{A}}^{2}v^{2}-4g_{x}^{2}u^{2}}
\approx\frac{-g_{\mathbf{A}}vM_{0}}{g_{x}^{2}u^{2}}\,,
\label{M2mxang}
\end{equation}
where $M_{0}=\frac{v}{2}\sqrt{g_{2}^{2}+g_{Y}^{2}}$ is again the original $Z$ boson mass in the SM.
The modified $Z$ boson and $Z^\prime$ boson masses are given by
\begin{align}
M_{Z} & \approx M_{0}(1-\frac{g_{\mathbf{A}}^{2}v^{2}}{8g_{x}^{2}u^{2}})\,,\label{ZMcoG}\\
M_{Z^\prime} & \approx g_x u (1 + \frac{g_{\mathbf{A}}^2 v^2}{ 2g_x^2 u^2})\,.
\end{align}
For the case that the dark fermion $\chi$ carries zero hypercharge,
the interactions of $Z$ boson with fermions are worked out as
\begin{align}
-\mathcal{L}_{Z} & =Z_{\mu}\big(R_{13}g_{x}J_{x}^{\mu}+R_{23}g_{Y}J_{Y}^{\mu}+R_{33}g_{2}J_{3}^{\mu}\big)  \label{GZcou} \\
 & =Z_{\mu} \big[ -\sin\psi g_{x}Q_{\chi}\bar{\chi}\gamma^{\mu}\chi -\sin\psi g_{x}
 (\mathbf{A} Y_i+\mathbf{B} q)\bar{f}_{i}\gamma^{\mu}f_{i}
 - \cos\psi s_W g_Y J_Y^\mu + \cos\psi c_W g_2 J_3^\mu \big]\,.  \nonumber
 %& =Z_{\mu} \big( -\sin\psi g_x J_x - \cos\psi s_W g_Y J_Y + \cos\psi c_W g_2 J_3 \big)\,.
\end{align}
The $Z^{\prime}$ interactions with fermions are given by
\begin{align}
-\mathcal{L}_{Z^{\prime}} & =Z_{\mu}^{\prime}\big(R_{11}g_{x}J_{x}^{\mu}+R_{21}g_{Y}J_{Y}^{\mu}+R_{31}g_{2}J_{3}^{\mu}\big) \label{GZpcou} \\
 & =Z_{\mu}^{\prime}\big[\cos\psi g_{x}Q_{\chi}\bar{\chi}\gamma^{\mu}\chi+\cos\psi g_{x}
 (\mathbf{A} Y_i+\mathbf{B} q)\bar{f}_{i}\gamma^{\mu}f_{i}
 - \sin\psi s_{W} g_{Y} J_{Y}^{\mu}+\sin\psi c_{W} g_{2} J_{3}^{\mu}\big]\,.  \nonumber
\end{align}
In Eqs.~(\ref{GZcou}) and (\ref{GZpcou}) the value of $Y_i$ depends on the left- and right-handedness of the SM fermions.

Writing $Z,Z^\prime$ interactions with the SM fermions in the standard form, we have
\begin{align}
    \mathcal{L}_{Z Z^\prime}=-\frac{e}{2s_W c_W}
    \big[ \bar{f}_i\gamma^\mu(v_i-a_i\gamma^5)f_i Z_\mu
    +\bar{f}_i\gamma^\mu (v_i^\prime-a_i^\prime\gamma^5)f_i Z^\prime_\mu
    \big]
\end{align}
where
\begin{align}
    v_i=& T_3^i \left(\cos\psi+\frac{\sin{\psi} g_x \mathbf{A}}{\sqrt{g_2^2 + g_Y^2}} \right)
    -2Q_i \left( s_W^2 \cos\psi +\frac{\sin{\psi} g_x \mathbf{A}}{\sqrt{g_2^2 + g_Y^2}} \right)
    +\frac{2\sin{\psi} g_x \mathbf{B} q_i}{\sqrt{g_2^2 + g_Y^2}} \, ,\label{M2Zvcou} \\
    a_i=& T_3^i \left(\cos\psi+\frac{\sin{\psi} g_x \mathbf{A}}{\sqrt{g_2^2 + g_Y^2}}\right)\,, \label{M2Zacou} \\
    v^\prime_i=& T_3^i\left(\sin{\psi}-\frac{\cos \psi g_x \mathbf{A}}{\sqrt{g_2^2 + g_Y^2}}\right)
    +2Q_i\left(\frac{\cos\psi g_x \mathbf{A}}{\sqrt{g_2^2 + g_Y^2}}-\sin{\psi} s_W^2\right)
    +\frac{2\cos\psi g_x \mathbf{B} q_i}{\sqrt{g_2^2 + g_Y^2}} \, ,\label{M2Zpvcou} \\
    a^\prime_i=& T_3^i \Big(\sin{\psi}-\frac{\cos \psi  g_x \mathbf{A}}{\sqrt{g_2^2 + g_Y^2}}\Big) \,,\label{M2Zpacou}
\end{align}
where $Q^i, T_3^i$ are respectively the electric charge and the third component of weak isospin
of the SM fermions;
$q_i$ is the $q$ quantum number carried by the SM fermions, and are set to zero in Section.~\ref{sec:DM},
although in a more general setup $q_i$ could be nonzero.

Looking into the expression of Eqs.~(\ref{M2Zvcou}) and (\ref{M2Zacou}),
we notice that the change to the original SM $Z$ boson coupling is proportional to $\sin \psi g_x \lesssim 10^{-4}$
which is a tiny modification.
We also notice that
although $g_{\mathbf{A}}= \mathbf{A} g_Y$ where $\mathbf{A}\sim \mathcal{O}(1)$,
can have a similar size to $g_Y$,
it only appears in the mixing angle $\psi \lesssim 10^{-3}$, c.f, Eq.~(\ref{M2mxang}),
and thus the value of $g_{\mathbf{A}}$ does not impose additional constraint to the model.

Here we would like to emphasize that this is the key of resolving the $W$ boson mass enhancement
using the $U(1)_{\mathbf{A} Y+\mathbf{B} q}$ model.
On the one hand, $W$ boson mass is enhanced through the mixing effect,
which is sensitive to the $g_{\mathbf{A}}$ value, c.f., Eq.~(\ref{WMco}) and Eq.~(\ref{M2Obpara});
on the other,
the modification to the couplings of $Z$ boson with SM fermions is proportional to $\sin \psi g_x$,
while the $Z^\prime$ to SM fermions couplings depend mainly on $g_x$,
thus with a proper choice of $g_x$ and $M_{Z^\prime}$ values,
all the current experimental constraints can be satisfied.

\section{Correction to the $W$ boson mass}\label{sec:WM}

The Peskin-Takeuchi oblique parameters $S, T, U$ are a set of three measurable quantities,
which parameterize new physics contributions to electroweak radiative corrections~\cite{Peskin:1990zt,Peskin:1991sw}.
Under the $S, T, U$ effective Lagrangian formulation~\cite{Holdom:1990xp,Burgess:1993vc},
new physics contributions are recast into effective operators using the original SM gauge fields,
giving rise to small shifts in $S, T, U$ parameters.
With the inclusion of an extra $U(1)$ gauge boson mixing with the SM gauge bosons,
$S, T, U$ parameters can be expressed by the changes in the redefinition of the gauge fields
as well as the mass shifts of the SM gauge bosons~\cite{Holdom:1990xp},
\begin{align}
	\alpha S &= 4 c_W s_W  (s_W^2 - c_W^2) \delta_{AZ} - 2 s_W c_W \delta_A + 2  s_W c_W \delta_Z\,,\\
    \alpha T &= 2 ( \delta_Z -  \tilde{\delta}_Z)\,,\\
    \alpha U &= -8 s_W^2 (s_W c_W \delta_{AZ} + s_W^2 \delta_A + c_W^2 \delta_Z)\,,
\end{align}
with $\delta_A,\delta_Z,\delta_{AZ}$ and $\tilde{\delta}_Z$ given by
\begin{align}
Z_{\rm SM} &= (1 + \delta_Z) Z\,, \\
A_{\rm SM} &= (1 + \delta_A) A + \delta_{AZ}  Z \,, \\
M_Z &= M_0 (1+\tilde{\delta}_Z) \,,
\end{align}
where $A_{\rm SM},Z_{\rm SM}$ are the original photon and $Z$ gauge fields from the SM,
whereas $A,Z$ are the physical photon and $Z$ boson in the final mass eigenbasis.

Under this formalism, values of all original SM parameters may alter with the presence of new physics.
All physical fields in the final mass eigenbasis as well as all altered physical quantities, %after rescaling or redefinition,
such as the electromagnetic fine-structure constant, take the observed experimental values~\cite{Burgess:1993vc}.
The shifts in $S, T, U$ can generate an enhancement of the $W$ boson mass~\cite{Peskin:1991sw,Burgess:1993vc},
\begin{equation}
	\Delta M_W^2 = \frac{ c_W^2 M_0^2}{c_W^2 - s_W^2} \left(
 -\frac{\alpha S}{2}  + c_W^2 \alpha T +\frac{c_W^2-s_W^2}{4s_W^2} \alpha U \right)\,.
 \label{WMco}
\end{equation}

\subsubsection*{Dark $U(1)_x$ extension of the SM}

For the dark $U(1)_x$ extension of the SM as described in Section.~\ref{sec:darkU1},
with some manipulation of Eqs.~(\ref{Trans33}) and (\ref{ZMco})
we find to the lowest order,
\begin{equation}
\delta_A =0\,,\quad
\delta_{AZ}= \delta^2 \epsilon^2 s_W c_W\,,\quad
\delta_Z = - \delta^2 \epsilon^2 s_W^2\,,\quad
\tilde{\delta}_Z = - \frac{1}{2} \delta^2 \epsilon^2 s_W^2\,,
\end{equation}
leading to
\begin{equation}
	\alpha S  = -\,4\,\delta^2 \epsilon^2 s_W^2  c_W^2\,,\quad
	\alpha T = - \,\delta^2 \epsilon^2 s_W^2 \,, \quad	\alpha U=0\,,
\end{equation}
where $\delta$ is the kinetic mixing parameter and $\epsilon = M_0 / M_{Z^\prime}$.
For the case that the heavy dark $U(1)_x$ gauge field %($M_1^2 \gg M_0^2$)
only mixes with the SM via gauge kinetic terms,
we notice that
both $S,T$ parameters have negative values, while the combination of the two generates an enhancement of the $W$ boson mass.
$U$ parameter is zero to this order and thus has no contribution.
The signs of $S,T$ may change after taking into account the mass mixing effect~\cite{Babu:1997st}.

\subsubsection*{General $U(1)_{\mathbf{A} Y+\mathbf{B} q}$ extension of the SM}

For the general $U(1)_{\mathbf{A} Y+\mathbf{B} q}$ extension of the SM as described in Section.~\ref{sec:generalU1},
with some manipulation of Eqs.~(\ref{Trans33G}) and (\ref{ZMcoG})
we find to the lowest order,
\begin{equation}
\delta_A =0\,,\quad \delta_{AZ}= 0\,,\quad
\delta_Z = -\frac{g_{\mathbf{A}}^2 v^2 M_0^2}{8g_x^4 u^4}\,,\quad
\tilde{\delta}_Z = - \frac{g_{\mathbf{A}}^2 v^2 }{ 8 g_x^2 u^2}\,,
\end{equation}
leading to
\begin{equation}
	\alpha S  = -\, \frac{s_W c_W g_{\mathbf{A}}^2 v^2 M_0^2 }{4 g_x^4 u^4}\,,\quad
	\alpha T = \, \frac{g_{\mathbf{A}}^2 v^2 }{4 g_x^2 u^2} \,, \quad	\alpha U=0\,,
\label{M2Obpara}
\end{equation}
where $g_{\mathbf{A}} = \mathbf{A} g_Y$, $g_x$ is the $U(1)_{\mathbf{A} Y+\mathbf{B} q}$ gauge coupling,
$v,u$ are the Higgs vev and extra scalar vev respectively.
For the case that the heavy $U(1)_{\mathbf{A} Y+\mathbf{B} q}$ gauge field
mixes with the SM through the gauge boson mass-square matrix,
$T$ offers positive and dominant contribution whereas $S$ takes the negative value,
and the combination of the two generates an enhancement of the $W$ boson mass.
$U$ parameter is still zero and thus has no contribution.

\subsubsection*{Experimental constraints and phenomenological implications}

Under the $S, T, U$ effective Lagrangian formulation,
all altered SM physical quantities take the observed experimental values,
considering the new physics effects.
Thus for the models we consider, the redefined $Z$ boson has the mass~\cite{ParticleDataGroup:2020ssz}
\begin{equation}
M_Z = 91.1876 \pm 0.0021~{\rm GeV}\,.
\end{equation}
We would like to emphasize here that very importantly, the $Z$ boson mass measurement error bar does not
correspond to the change to the original SM $Z$ boson mass, shown in Eq.~(\ref{ZMco}) and Eq.~(\ref{ZMcoG}).
In both of the models we discuss, the measured $Z$ boson mass from experiment is identified to be the modified $Z$ boson after the $U(1)$ mixing
in the final mass eigenbasis.

Current experiments set strong constraints on the mass and couplings of
the extra $U(1)$ gauge bosons.
%LEP-II and LHC both set stringent bounds on an extra $Z^\prime$ boson.
Constraints from $e^+e^-$ colliders are from resonance production of $e^+e^-\to \ell^+ \ell^-$ processes.
At the LHC, $Z^\prime$ boson can be detected through Drell-Yan processes or by examining dijet resonances.
Stringent bound has been set on an extra massive $Z^\prime$ gauge boson most recently by ATLAS~\cite{ATLAS:2019erb}.
Using ATLAS and CMS data,
a model-independent study~\cite{Fairbairn:2016iuf} is carried out and
the upper limit on the couplings of $Z^\prime$ to quarks is given,
corresponding to a set of $M_{Z^\prime}$ and $\Gamma_{Z^\prime}$ values.
%In particular, for the case that $\Gamma_{Z^\prime}/M_{Z^\prime} \lesssim 0.01$,
%the upper bound on $Z^\prime$ to quark coupling is 0.1~\cite{Fairbairn:2016iuf}.
The $Z^\prime$ couplings to quarks of both our models shown in Eqs.~(\ref{M1Zpvcou}) and (\ref{M1Zpacou}),
Eqs.~(\ref{M2Zpvcou}) and (\ref{M2Zpacou}), are well below the limits presented in~\cite{Fairbairn:2016iuf}.

The limit of models with an extra $U(1)$ with coupling $g_{Z^\prime \bar{\ell} \ell}$ to SM leptons,
can be written as
\begin{equation}
\frac{M_{Z^\prime}}{g_{Z^\prime \bar{\ell} \ell} } \gtrsim 12~{\rm TeV}\,,
\end{equation}
which is a stringent constraint for $U(1)$ dark sectors connecting with the SM through mixing effects.
All benchmark points in the figures presented in this paper have taken into account the above constraints.

%The Fermilab E989 experiment has measured $a_\mu = (g-2)_\mu/2$ with a very high accuracy that
%\begin{equation}
%a_\mu^{\rm E989} = 116592040(54)\times 10^{-11}\,.
%\end{equation}
%Comparing with the Standard Model result XXX
%\begin{equation}
%a_\mu^{\rm SM} =  116591810(43)\times 10^{-11}\,.
%\end{equation}
%Though this discrepancy does not reach $5\sigma$ to become a discovery of new physics,
%this combined result seems to be more convincing and may suggest new physics contributions.

The combined results~\cite{ParticleDataGroup:2018ovx} from Fermilab E989~\cite{Muong-2:2021ojo}
and Brookhaven E821~\cite{Muong-2:2006rrc} experiments show
a $4.2\sigma$ deviation from the SM prediction~\cite{Aoyama:2020ynm}, which is
\begin{equation}
\Delta a_\mu =  251(59) \times 10^{-11}\,.
\end{equation}
This combined result may suggest new physics contributions.
The $Z^\prime$ contribution to muon $g-2$ is given by
\begin{equation}
	\Delta (g_\mu -2) \approx \frac{g_{Z^\prime \bar{\mu} \mu}^2 m_\mu^2}{6 \pi^2 M_{Z^\prime}^2}\,.
\end{equation}
However, for a TeV scale $Z^\prime$ gauge boson with a very weak coupling strength to muon pairs,
this contribution is negligible.

We also investigate the implications of the two models we discussed in the previous sections
on the precisely determined observables in the electroweak sector.
Especially we find the modified $Z$ decay widths of both models are within the error bar of the current measurements,
and the details analysis are given in Appendix~\ref{sec:AppA}.

\begin{figure}[!t]
\center{\includegraphics[width=0.6\textwidth]{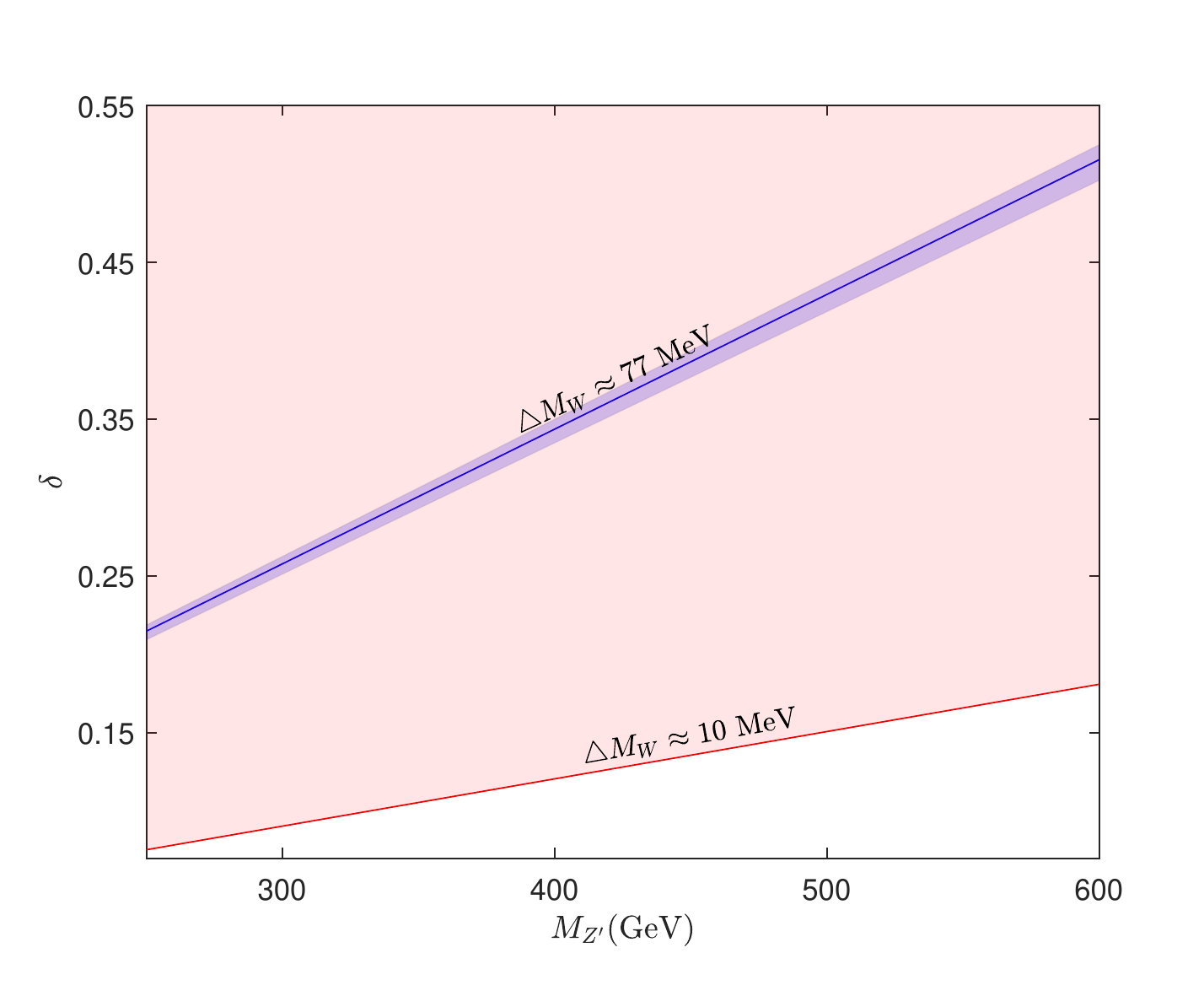}}
\caption{(color online) An exhibition of the $W$ boson mass enhancement
  generated by the dark $U(1)_x$ extension of the SM with the $U(1)_x$ gauge field
  mixed with the SM through gauge kinetic terms.
  $\delta$ is the kinetic mixing parameter.
  The dark blue line corresponds to the central value of the $W$ boson mass enhancement $\Delta M_W\sim77$~MeV,
  and the light blue region took into account the experimental error bar and theoretical uncertainties.
  The region in red is ruled out by experimental constraints.
  The lower boundary of the red region shows the upper limit of $\delta$,
  corresponding to 10~MeV enhancement of the $W$ boson mass.}
\label{fig:WMK}
\end{figure}

%The correction to the $W$ boson mass is given by Eq.~(\ref{WMco}).

For the dark $U(1)_x$ extension of the SM with kinetic mixing,
the $\delta-M_{Z^\prime}$ plot is shown in Fig.~\ref{fig:WMK}.
The blue region can explain the $W$ boson mass enhancement
considering the experimental error bar and theoretical uncertainties.
The dark blue line corresponds to the central value of the $W$ boson mass enhancement $\Delta M_W\sim77$~MeV.
To achieve 77~MeV enhancement,
$M_{Z^\prime}$ can only stay in the region of several hundred GeV with a rather large $\mathcal{O}(1)$ kinetic mixing.
Such kinetic mixing will generate $Z^\prime$ to SM fermions coupling strengths $\sim \delta g_Y$,
c.f., Eq.~(\ref{ZPSMcoupK}) for ${Z^\prime \bar{f} f}$ couplings.
Thus the entire red region is ruled out by experimental constraints as mentioned above.
The reason is simple.
For the dark $U(1)_x$ extension of the SM,
the kinetic mixing generates both the $W$ boson mass enhancement and the $Z-Z^\prime$ mixing.
Achieving the desired value of the $W$ boson mass enhancement
requires a $\mathcal{O}(1)$ kinetic mixing,
rendering a rather large $Z-Z^\prime$ mixing.
As a result,
$Z^\prime \bar{f}f$ couplings are comparable with $Z\bar{f}f$ couplings,
and thus such $Z^\prime $ faces severe restrictions from experiments.
With the current experimental bounds,
the dark $U(1)_x$ extension of the SM with only kinetic mixing can offer at most 10~MeV
enhancement to the $W$ boson mass,
corresponding to the $\delta$ upper limit line as shown in Fig.~\ref{fig:WMK}.

\begin{figure}[!t]
\center{\includegraphics[width=0.6\textwidth]{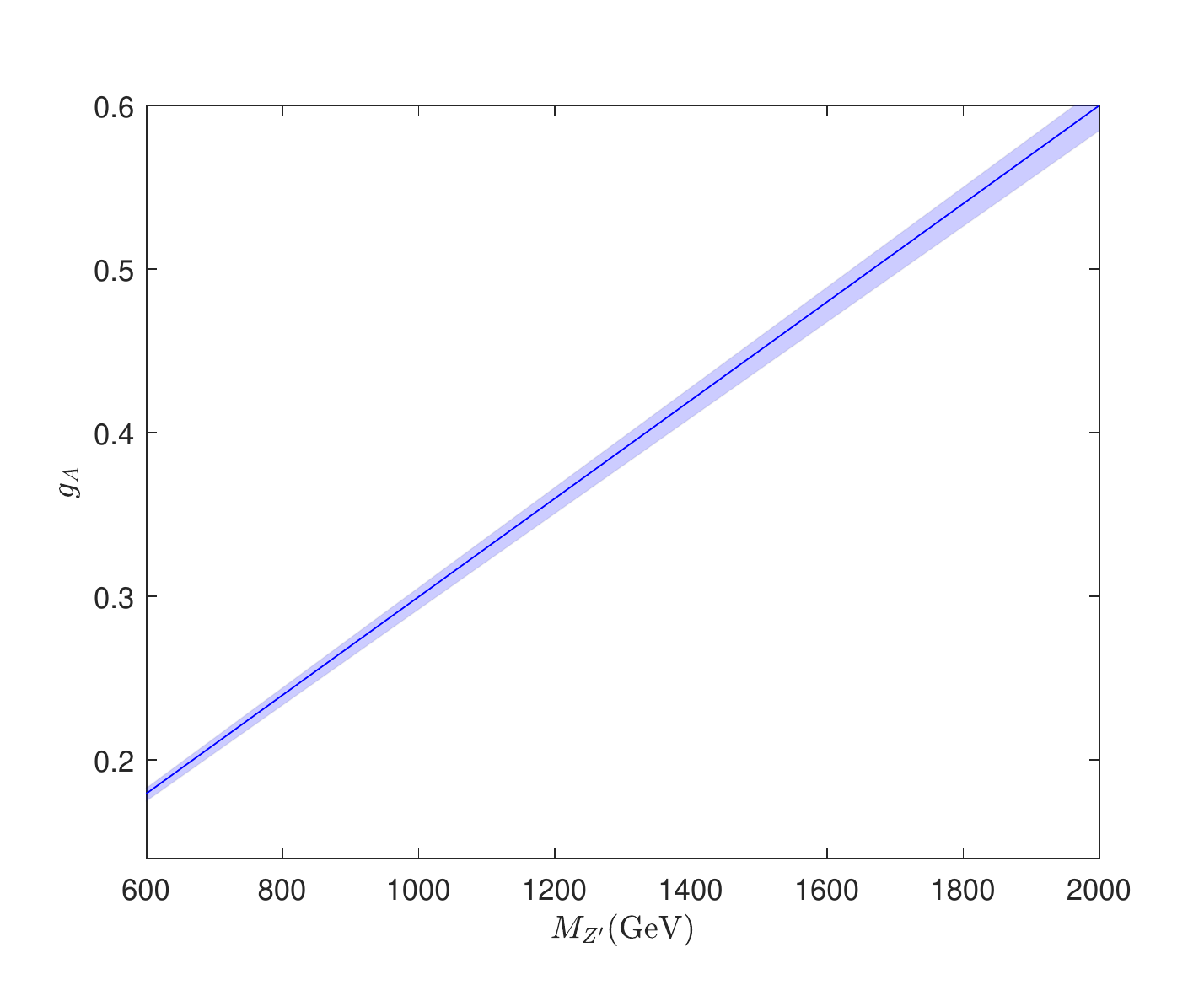}}
\caption{(color online) An exhibition of the $W$ boson mass enhancement from
  $U(1)_{\mathbf{A} Y+\mathbf{B} q}$ extension of the SM.
  The dark blue line corresponds to the central value of the $W$ boson mass enhancement $\Delta M_W\sim77$~MeV,
  and the light blue region took into account the experimental error bar and theoretical uncertainties.
  The entire blue region can satisfy all the current experiential constraints
  with a proper choice of $g_x$ and $M_{Z^\prime}$.
  %All experimental constraints are already taken into account.
  }
\label{fig:WMG}
\end{figure}

We now turn to the general $U(1)_{\mathbf{A} Y+\mathbf{B} q}$ extension of the SM,
the $g_{\mathbf{A}} - M_{Z^\prime}$ plot is shown in Fig.~\ref{fig:WMG}.
The blue region can well explain the $W$ boson mass enhancement
considering the experimental error bar and theoretical uncertainties.
The dark blue line corresponds to the central value of the $W$ boson mass enhancement of 77~MeV.
We already mentioned in the end of Section.~\ref{sec:generalU1},
for this case the mixing is generated via the $\mathbf{A} Y$ part of the gauged $U(1)$,
thus one has the freedom to tune the $U(1)$ gauge coupling $g_x$ as well as the $Z^\prime$ mass.
As a consequence,
various experimental constraints are not difficult to satisfy for the $U(1)_{\mathbf{A} Y+\mathbf{B} q}$ extension of the SM.
The entire blue region in Fig.~\ref{fig:WMG} can pass all the current experiential constraints
with a proper choice of $g_x$ and $Z^\prime$ mass, which is fairly easy to achieve.

\section{Dark matter candidate from $U(1)$ dark sector}\label{sec:DM}

Despite the great success of the SM, the nature of dark matter remains to be a puzzle of particle physics and cosmology.
The $U(1)$ dark sector includes dark fermions only charged under the extra $U(1)$ gauge group,
which are stable and thus the natural dark matter candidates.
We now investigate the dark fermion as dark matter candidate in the $U(1)_{\mathbf{A} Y+\mathbf{B} q}$ extension of the SM.
The dark fermion part of the Lagrangian is given by Eq.~(\ref{DF}),
and we consider the dark fermion $\chi$ carries the $U(1)$ charge $Q_\chi = +1$.
Here we assume the dark fermions carry zero hypercharge,
and thus they couple to the $Z$ boson only through the mass mixing effect.
We also assume the SM fermions carry no $q$ quantum number for simplicity,
thus they couple to $Z^\prime$ only through $\mathbf{A}  Y$ and the mass mixing.

The mass mixing between the $U(1)_{\mathbf{A} Y+\mathbf{B} q}$ and the SM mediates interactions between
dark fermions and SM particles, which can be an efficient mechanism to reduce the
dark fermion primordial density and to achieve the current observed value of the dark matter relic density via freeze-out process.
The dark fermion $\chi, \bar{\chi}$ can annihilate\footnote{
Dark fermion $\chi$ can also annihilate to $W^+W^-$ final state through $Z$ pole, with
\begin{equation}
\langle\sigma v\rangle_{W W} \simeq \frac{g_2^4 \tan \theta_W}{16 \pi M_W^2}\Big[\left|V_\chi\right|^2\big(1-\frac{v^2}{6}\big)+\left|A_\chi\right|^2 \frac{v^2}{3}\Big] \frac{m_\chi^2}{m_W^2}\,,
\end{equation}
where $V_\chi = -2\sin\psi g_x Q_\chi$ and $A_\chi=0$ in the $U(1)_{\mathbf{A} Y+\mathbf{B} q}$ model, c.f., Eq.~(\ref{GZcou}).
We find this contribution is 2 orders lower than the $\chi\bar{\chi}$ annihilation to SM fermions.
$\bar{\chi}\chi\to WW$ can also occur via the $Z^\prime$ pole,
which is another 2 orders lower
compared to the $Z$ pole contribution, and is thus dropped.
We also comment here that  $\bar{\chi}\chi\to Zh$ process
which only depends on $A_\chi$ is absent for our model.
} through the $Z$ and $Z^\prime$ poles to pairs of SM fermions $f_i$, i.e.,
\begin{equation}
\chi+\bar{\chi} \xrightarrow{\quad Z,\, Z^\prime\quad } f_i + \bar{f}_i\,. \label{DMann}
\end{equation}
Using the vector and axial couplings carried out in Eqs.~(\ref{M2Zvcou})-(\ref{M2Zpacou}),
the cross-section of these processes are given by
\begin{align}
\sigma_{\bar{\chi}\chi \to \bar{f}_i f_i} (s)
&=\frac{N_c g_2^2 g_x^2 Q_\chi^2 }{12 \pi c_W^2 s }  (s+2m_\chi^2) \sqrt{ \frac{s-4m_i^2}{s-4 m_\chi^2} } \times \\
&\quad \Big\{
\frac{\sin^2\psi  \big[ v_i^2 (s+2m_i^2) + a_i^2 (s-4m_i^2) \big]}
{ (s-M_Z^2)^2 + M_Z^2 \Gamma_Z^2 }
+\frac{ v^{\prime 2}_i (s+2m_i^2) + a^{\prime 2}_i (s-4m_i^2)}
{(s-M_{Z^\prime}^2)^2 + M_{Z^\prime}^2 \Gamma_{Z^\prime}^2}  \nonumber  \\
&\quad -\frac{2 \sin\psi  \big[v_i v_i^\prime (s + 2m_i^2) + a_i a_i^\prime (s-4m_i^2)\big]}
{ \big[(s-M_Z^2)^2 + M_Z^2 \Gamma_Z^2\big] \big[(s-M_{Z^\prime}^2)^2 + M_{Z^\prime}^2 \Gamma_{Z^\prime}^2\big]}
\mathcal{F}(s) \Big\}\,,\nonumber
\end{align}
where $f_i$ are SM fermions with masses $m_i$, $N_c$ is the color factor, $m_\chi$ is the dark matter mass,
and the form factor is given by
\begin{equation}
\mathcal{F}(s)=(s-M_Z^2)(s-M_{Z^\prime}^2) + \Gamma_Z\Gamma_{Z^\prime} M_Z M_{Z^\prime}\,.
\end{equation}
The modified $Z$ boson decay width is carried out in Appendix \ref{sec:AppA},
while the $Z^\prime$ decay width can be calculated as
\begin{equation}
\Gamma_{Z^\prime} = \Gamma_{Z^\prime \to \bar{\chi} \chi} + \sum_i \Gamma_{Z^\prime \to \bar{f}_i f_i} + \Gamma_{Z^\prime \to W^+W^-}\,,
\end{equation}
where
\begin{align}
\Gamma_{Z^\prime \to \bar{\chi} \chi} &= \frac{(g_x Q_\chi)^2 M_{Z^\prime}}{12\pi}
\sqrt{1-\frac{4m_\chi^2}{M_{Z^\prime}^2}} \Big(1+ \frac{2 m_\chi^2}{M_{Z^\prime}^2}\Big)\,,\\
	\Gamma_{Z^\prime\to \bar{f_i} f_i}&=\frac{N_c e^2 M_{Z^\prime} }{48 \pi s_W^2c_W^2} \sqrt{1-\frac{4 m_{i}^{2}}{M_{Z^\prime}^{2}}}\left[v_{i}^{\prime2}\left(1+\frac{2 m_{i}^{2}}{M_{Z^\prime}^{2}}\right)+a_{i}^{\prime2}\left(1-\frac{4 m_{i}^{2}}{M_{Z^\prime}^{2}}\right)\right],\\
	\Gamma_{Z^\prime\to W^+W^-}&=\frac{g_{Z W W}^{2} M_{Z^\prime}}{192 \pi} \sin^2\psi\left(\frac{M_{Z^\prime}}{M_{Z}}\right)^{4}\left(1-4 \frac{M_{W}^{2}}{M_{Z^\prime}^{2}}\right)^{\frac{3}{2}}\left(1+20 \frac{M_{W}^{2}}{M_{Z^\prime}^{2}}+12 \frac{M_{W}^{4}}{M_{Z^\prime}^{4}}\right).
\end{align}

With the processes Eq.~(\ref{DMann}), dark fermions annihilate to SM fermion pairs and freeze out as the Universe cools down.
The dark matter relic density can be computed as
%The current relic density: $ \Omega h^{2} \approx 0.1199 \pm 0.0027$.
\begin{equation}
		\Omega h^{2} \approx \frac{2 \times 1.07 \times 10^{9} \mathrm{GeV}^{-1}}{\sqrt{g_{*}} M_{\mathrm{Pl}} J\left(x_{f}\right)}\,,
\quad{\rm with}\
J\left(x_{f}\right)=\int_{x_{f}}^{\infty} \frac{\langle\sigma v\rangle}{x^{2}} \mathrm{~d} x\,,
\end{equation}
where $x_f$ is the dark matter mass over the freeze-out temperature,
defined as at $x_f$, $Y - Y_{\rm EQ} \sim cY_{\rm EQ}$ and $c$ is an $\mathcal{O}(1)$ value.
The thermal averaging cross-section is given by
\begin{equation}
\langle\sigma v\rangle=\frac{\int_{4 m_\chi^{2}}^{\infty} \mathrm{d} s \,s \sqrt{s-4 m_\chi^{2}} K_{1}(\sqrt{s} / T) \sigma v}{16 T m_\chi^{4} K_{2}^{2}(m_\chi/T)}\,.
\end{equation}
For a heavy $Z^\prime$ gauge boson with $\mathcal{O}({\rm TeV})$ mass,
we consider the dark fermions annihilate via a narrow Breit-Wigner $Z^\prime$ resonance
which can generate a large enough annihilation
cross-section in accordance with the observed dark matter relic abundance,
and thus the mass of dark matter is around half of the $Z^\prime$ mass.

\begin{figure}[!h]
\center{\includegraphics[width=0.6\textwidth]{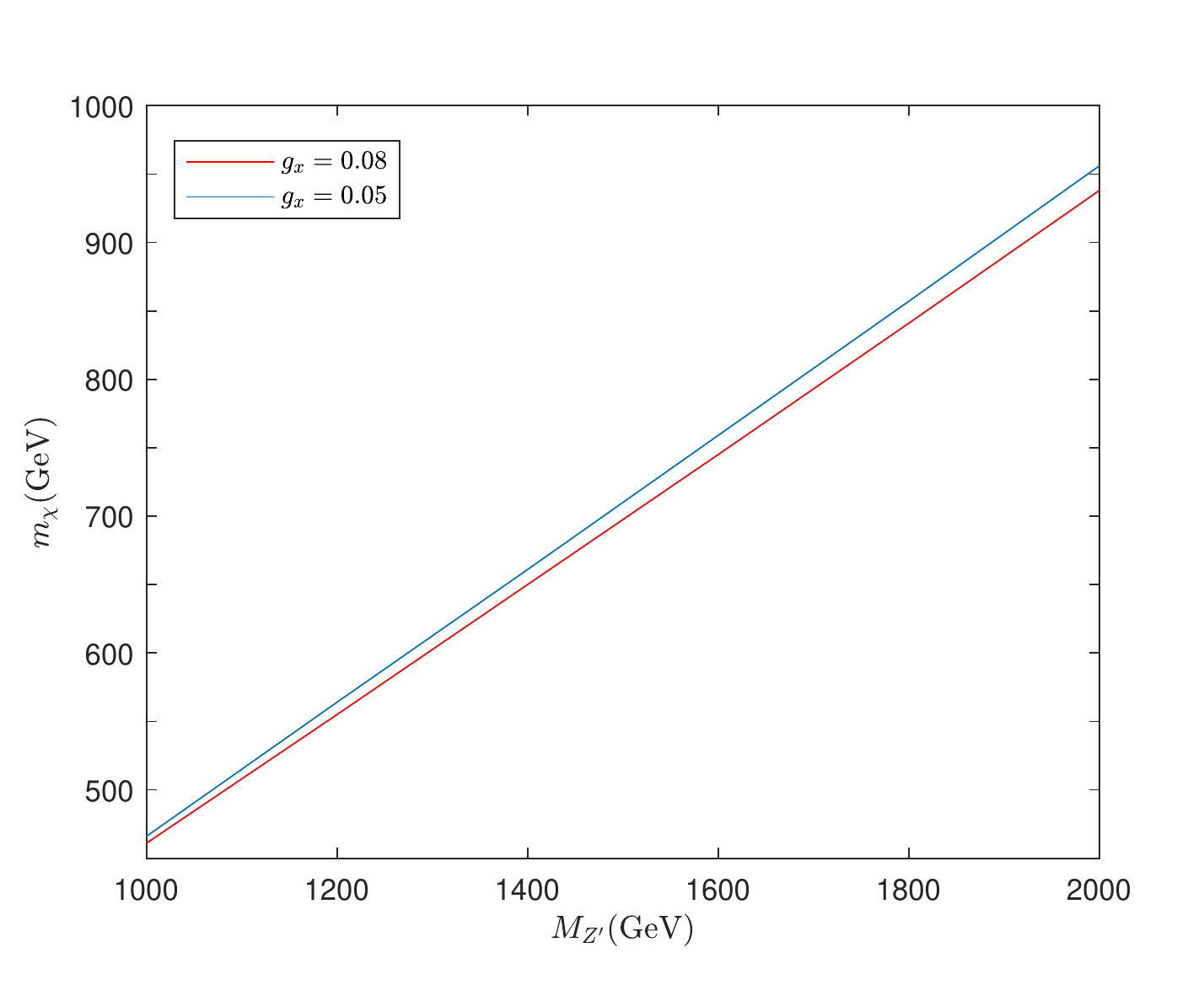}}
\caption{(color online) An exhibition of the dark matter mass verses the mass of $Z^\prime$ with two different dark sector gauge couplings.
  Data points on the lines present the observed value of dark matter relic density.}
\label{fig:DMRD}
\end{figure}

\begin{figure}[!h]
\center{\includegraphics[width=0.7\textwidth]{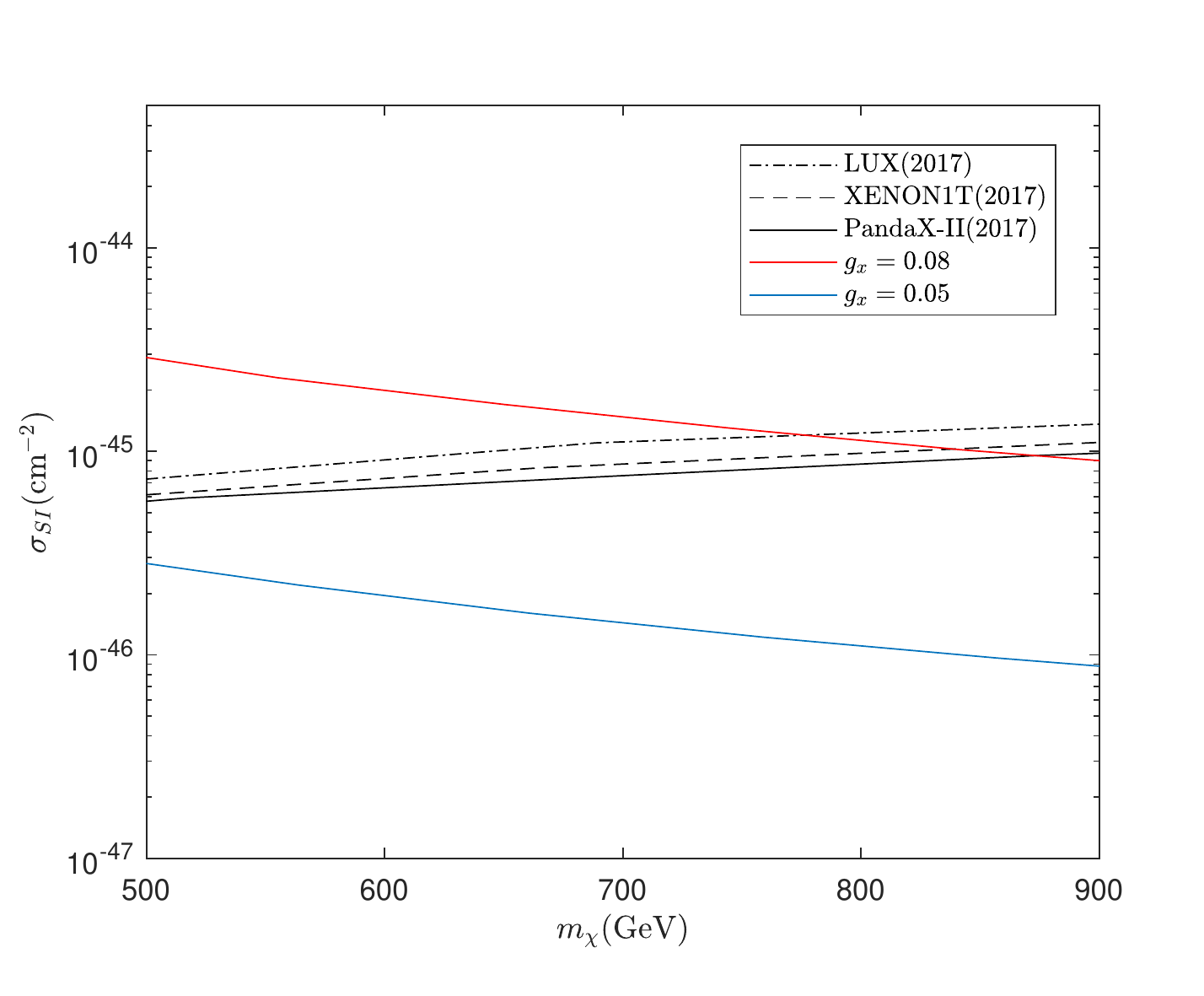}}
\caption{(color online) A display of the current constraints from dark matter direct detection experiments
  as well as benchmark points of dark matter from the $U(1)_{\mathbf{A} Y+\mathbf{B} q}$ extension of the SM.
  Data points above the black lines showing the bounds of direct detection experiments
  are excluded.}
\label{fig:DD}
\end{figure}

The mass of the corresponding dark matter candidate is shown in Fig.~\ref{fig:DMRD},
with two different values of dark sector gauge couplings $g_x$.
To accommodate the observed dark matter relic abundance using Wigner enhancement effect,
the mass of the dark matter is around half of the $Z^\prime$ mass.
As $M_{Z^\prime}$ increases and $g_x$ decreases,
the mass of dark matter approaches more to the half of ${Z^\prime}$ mass
to achieve the observed dark matter relic density.
The most stringent constraint on the model is coming from dark matter direct detection experiments.
We plot the spin-independent cross-section verses dark matter mass in Fig.~\ref{fig:DD}.
We find for the model we discuss,
data points above the black lines are excluded by the current direct detection experiments~\cite{LUX:2016ggv,XENON:2017vdw,PandaX-II:2017hlx}.
For example, for $U(1)$ gauge coupling $g_x=0.08$,
data points with dark matter mass less than $\sim900$~GeV are excluded by experiments.
While data points with $g_x=0.05$
are still viable dark matter candidates and
will be in reach of the next generation of dark matter direct detection experiments.
Improved experiments in the future for the large mass region of dark matter with better sensitivities should be able to test the model.

\begin{figure}[!h]
\begin{center}
\includegraphics[scale=0.36]{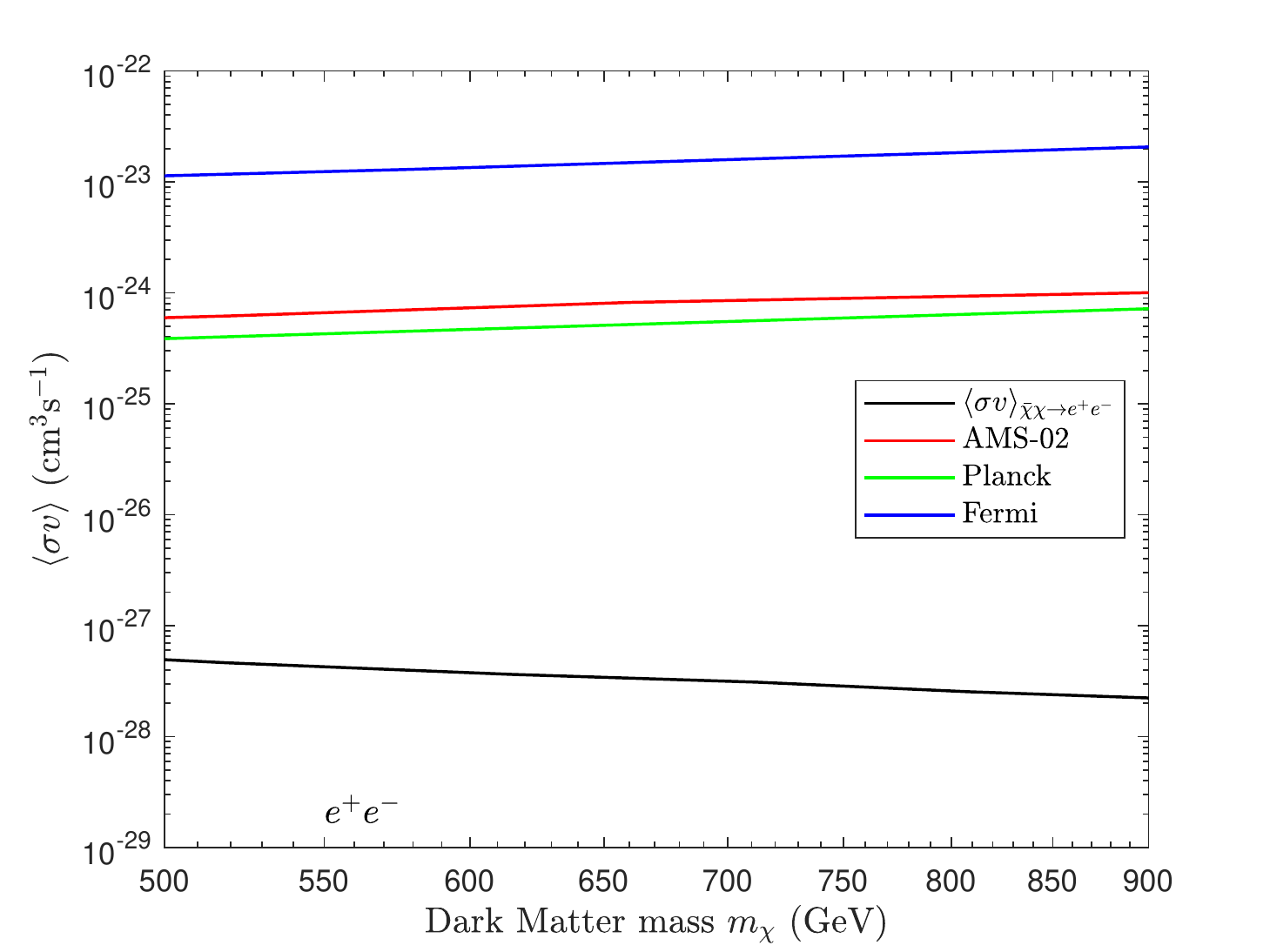}
\includegraphics[scale=0.36]{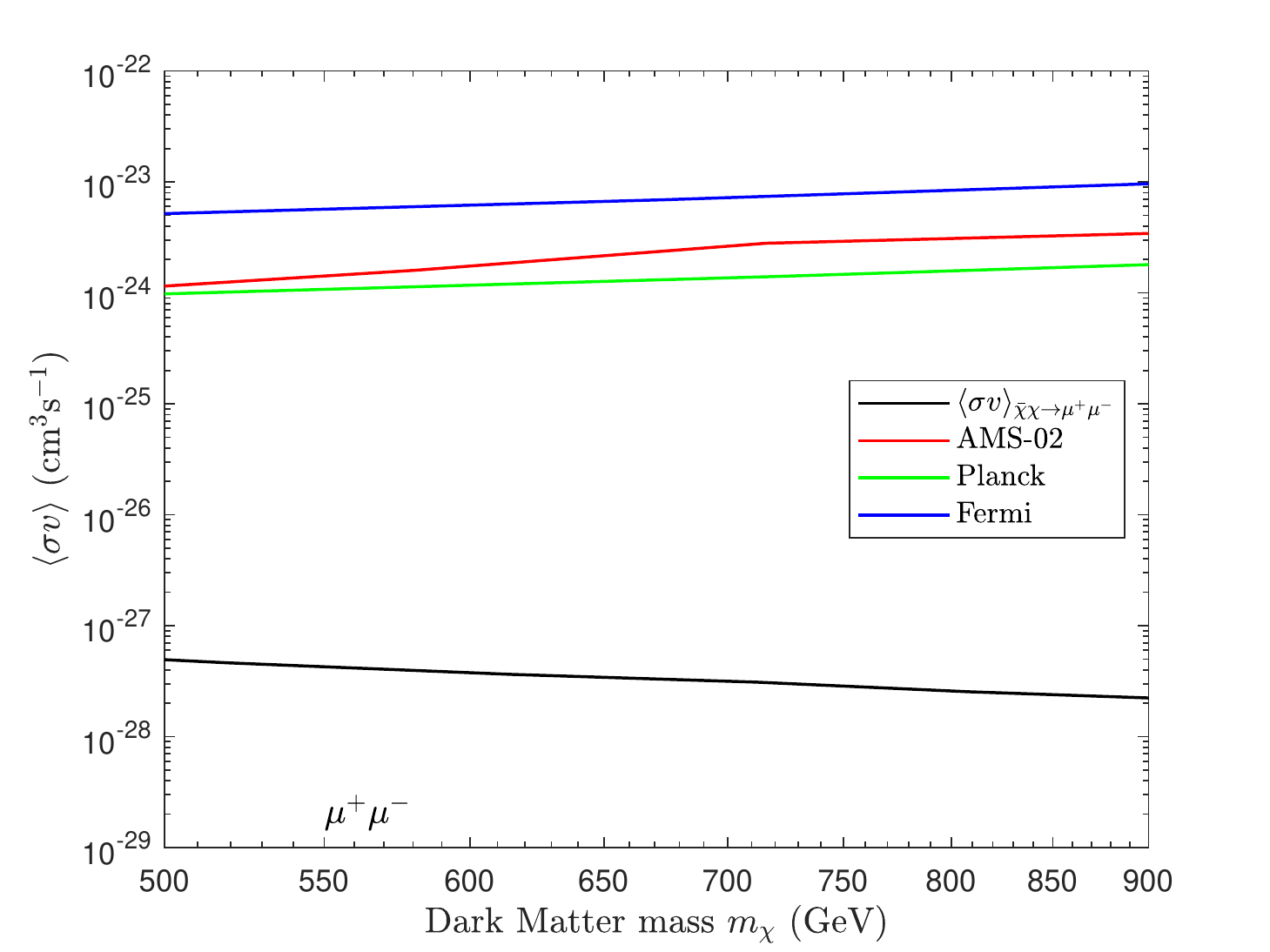}
\includegraphics[scale=0.36]{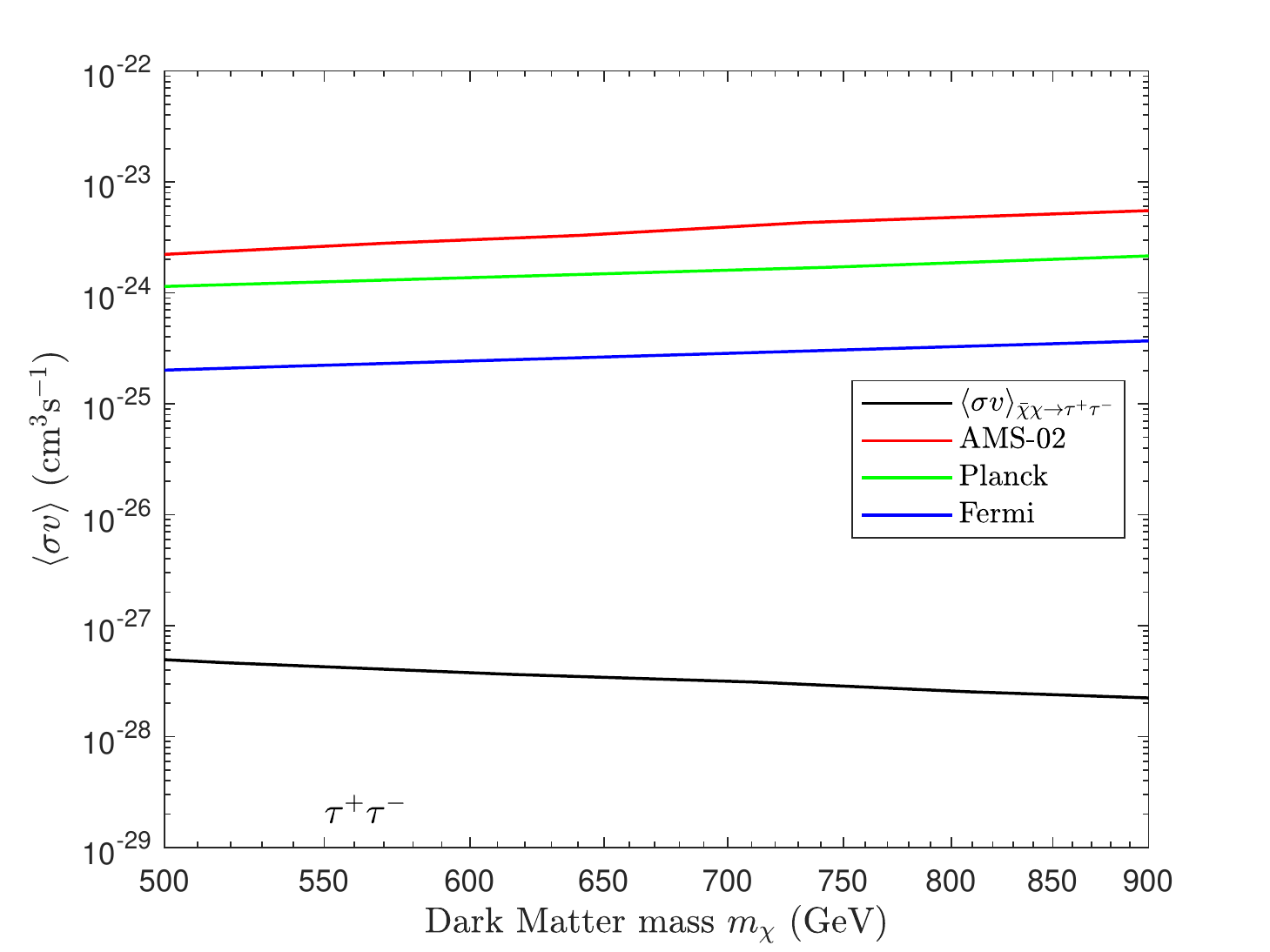}
\\
\includegraphics[scale=0.36]{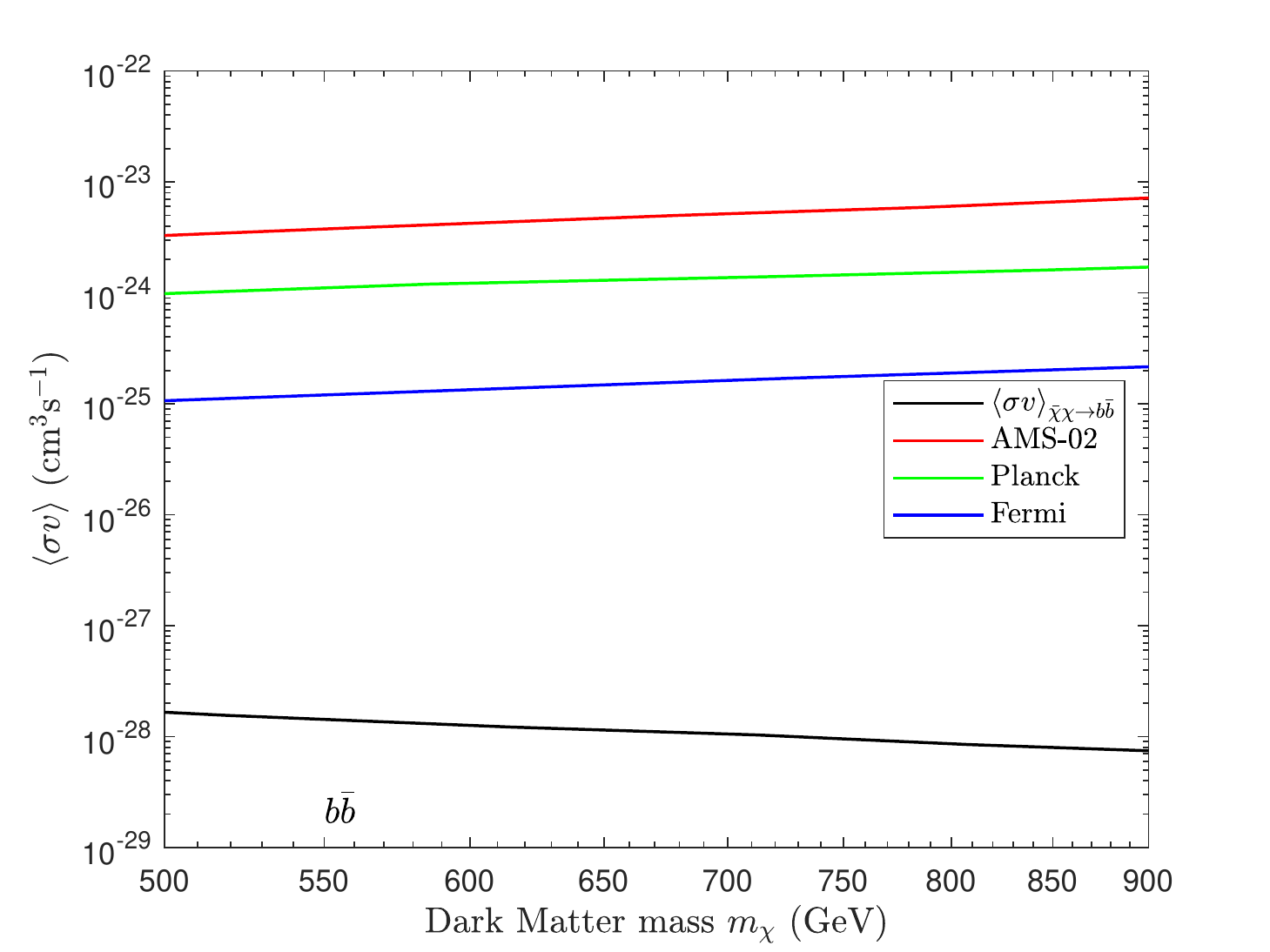}
\includegraphics[scale=0.36]{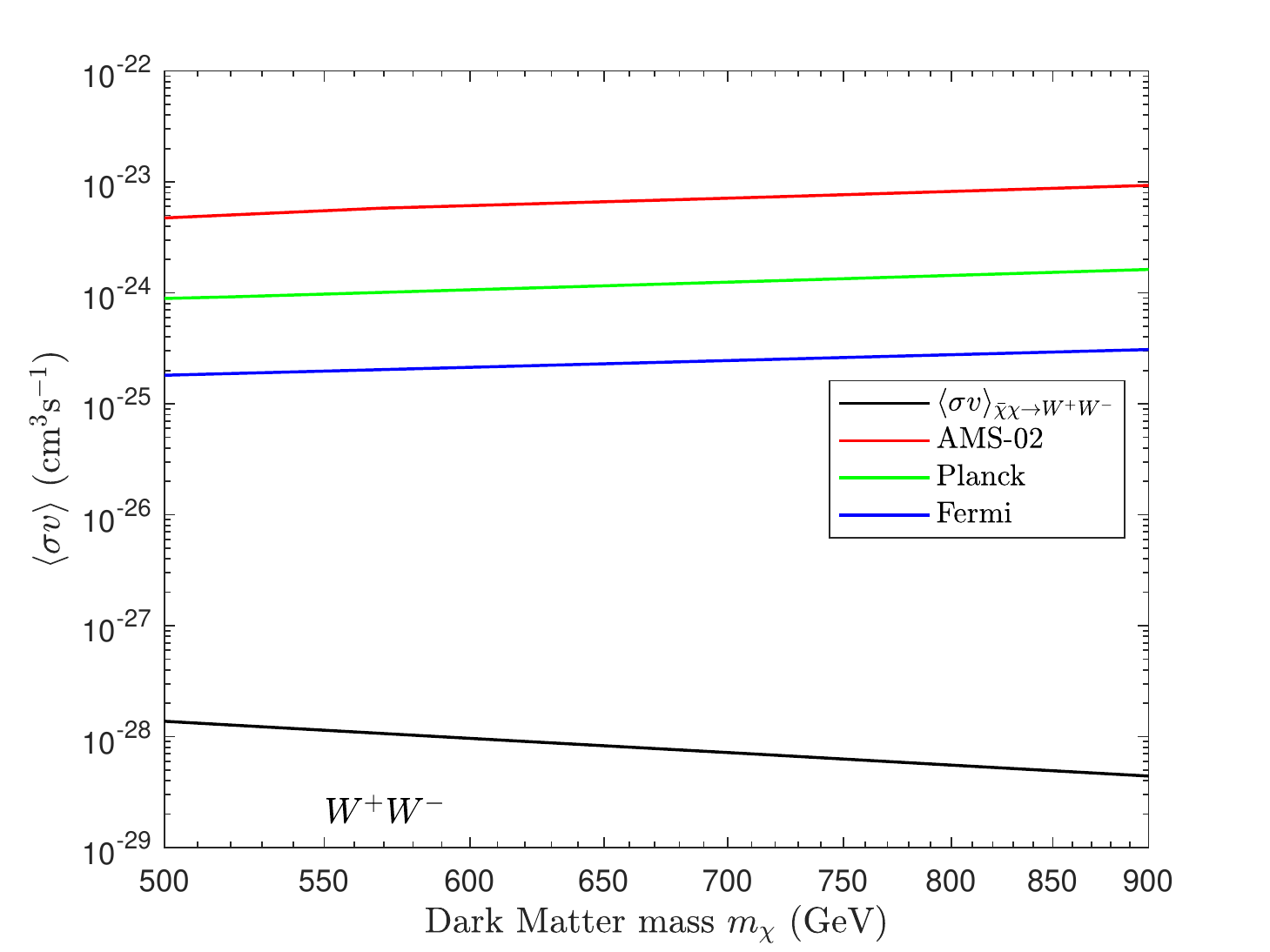}
\end{center}
\caption{(color online) A display of current constraints from dark matter indirect detection experiments
  as well as benchmark points of dark matter from the $U(1)_{\mathbf{A} Y+\mathbf{B} q}$ extension of the SM with $g_x = 0.08$.
  All benchmark points are below the current constraints.}
\label{fig:ID}
\end{figure}

For stable dark matter, the indirect detection bound comes mostly from dark matter annihilation to SM particles.
In Fig.~\ref{fig:ID},
we plot the constraints from CMB (green)~\cite{Planck:2015fie}, AMS-02 (red)~\cite{AMS:2014xys,AMS:2014bun}
and {\it Fermi} (blue)~\cite{Fermi-LAT:2015att} for various SM final states.
We take the larger $g_x=0.08$ value used in dark matter analysis Fig.~\ref{fig:DMRD} and Fig.~\ref{fig:DD},
which gives a larger annihilation cross section.
Other indirect detection bounds also includes dark matter annihilation to $\gamma\gamma,gg,hh$ final states,
which are absent in our model.

\section{Conclusion}\label{sec:Con}

Recently the CDF collaboration has announced a more precise measurement of the $W$ boson mass
with a central value significantly larger than the SM prediction about 77~MeV with $7\sigma$ deviation.
This result is in favor of the presence of new physics beyond the SM.
It is thus crucial to explore phenomenological implications of the new CDF result on the $W$ boson mass measurement.
In this paper,
we perform a comprehensive study on two extra $U(1)$ extensions of the SM.
One is the dark $U(1)_x$ extension, the other is the $U(1)_{\mathbf{A} Y+\mathbf{B} q}$ extension.
The mixing effects between the extra $U(1)$ dark sector and the SM are discussed in detail.
We find for the scenario that the dark $U(1)_x$ has only kinetic mixing with the SM,
the $W$ boson mass can be enhanced by at most 10~MeV according to various experimental constraints.
While we demonstrate for the $U(1)_{\mathbf{A} Y+\mathbf{B} q}$ extension of the SM,
one can achieve a consistent scenario in agreement with all existing experimental results,
and explain the $W$ boson mass enhancement $\sim77$~MeV as well as the nature of dark matter.
The mixing mediates interactions between dark fermions and SM particles
through the exchange of $Z$ boson as well as the $\mathcal{O}({\rm TeV})$ $Z^\prime$ gauge boson.
The dark fermions can annihilate through $Z,Z^\prime$ poles and become relic in the Universe
and are thus natural dark matter candidates.
We calculate the dark matter relic density and fit our model points with dark matter direct
and indirect detection experimental bounds.
The viable benchmark points of the model offer dark matter candidates
within the reach of future dark matter direct detection experiments. \\

\noindent
\textbf{Acknowledgments: }
WZF is supported in part by the National Natural Science Foundation of China under Grant No. 11905158 and No. 11935009,
and Natural Science Foundation of Tianjin City under Grant No. 20JCQNJC02030.

\appendix

\section{Electroweak precision test of $Z$ boson}\label{sec:AppA}

We now investigate the implications of the analysis in previous sections on the precisely determined observables in the electroweak sector.
One needs to make sure the modified $Z$ boson, now carrying a tiny fraction of $Z^\prime$, passes the electroweak precision test.
The couplings of the $Z$ boson to SM fermions are elevated from the tree level expressions
\begin{align}
    v_i=&\sqrt{\rho_i}\, \big(T_i^3 C_1 + 2\kappa_i Q_i C_2\big) \,,\\
    a_i=&\sqrt{\rho_i}\, T_i^3 C_1\,,
\end{align}
where $\rho_i,\kappa_i$ are given in~\cite{Feldman:2006wb}.
For the pure kinetic mixing $U(1)$ model, the two parameter $C_1,C_2$ are
\begin{align}
    C_1^{\rm km}=&\cos\psi +s_\delta s_W \sin\psi \,,\\
    C_2^{\rm km}=&- s_W^2 (\cos\psi +s_\delta s_W^{-1} \sin\psi) \,,
\end{align}
while for the $U(1)_{{\mathbf{A} Y + \mathbf{B} q}}$ model
\begin{align}
    C_1^{{\mathbf{A} Y + \mathbf{B} q}}=&\cos\psi+\sin\psi  g_x A \big/ \sqrt{g_2^2 + g_Y^2} \,,\\
    C_2^{{\mathbf{A} Y + \mathbf{B} q}}=&-\sin\psi g_x A -\cos\psi s_W g_Y\,.
\end{align}
The decay of the $Z$ boson into lepton anti-lepton and
quark anti-quark pairs (excluding the top) in the on-shell renormalization scheme
is given by~\cite{Baur:2001ze,Feldman:2006wb}
\begin{align}
\Gamma(Z \rightarrow f_i \bar{f_i}) &=N_f^c \mathcal{R}_i \Gamma_o \sqrt{1-4 \mu_i^2}\left[\left|v_i\right|^2 (1+2 \mu_i^2)+\left|a_i\right|^2 (1-4 \mu_i^2)\right]\,, \\
\mathcal{R}_i &=\left(1+\delta_f^{Q E D}\right)\left(1+\frac{N_i^c-1}{2} \delta_f^{Q C D}\right)\,, \\
\delta_i^{Q E D} &=\frac{3 \alpha}{4 \pi} Q_i^2\,, \\
\delta_i^{Q C D} &=\frac{\alpha_s}{\pi}+1.409\left(\frac{\alpha_s}{\pi}\right)^2-12.77\left(\frac{\alpha_s}{\pi}\right)^3-Q_i^2 \frac{\alpha \alpha_s}{4 \pi^2}\, .
\end{align}
where $\mu_i=m_i/M_Z$, $\Gamma_o=G_F M_Z^3 / 6 \sqrt{2} \pi$.

For the kinetic mixing $U(1)$ model we show the modified $Z$ boson decay width of two benchmark points
are taken from the two edges of the red lower bound from Fig.~\ref{fig:WMK},
which could only enhance $W$ boson mass about 10~MeV.
We find
\begin{align}
     \Gamma_Z(\delta=0.18,M_1=600~\text{GeV})\approx&\,2.4934~\text{GeV},
     \big |{\Gamma_Z^{\rm km}-\Gamma_Z^{EX}}\big|\approx 0.77~\text{errorbar}\, ,\\
       \Gamma_Z(\delta=0.09,M_1=300~\text{GeV})\approx&\,2.4934~\text{GeV},
       \big|\Gamma_Z^{\rm km}-\Gamma_Z^{EX}\big|\approx 0.77~\text{errorbar}\, .
\end{align}
All other points from Fig.~\ref{fig:WMK} also fall into the experimental error bars of the $Z$ boson decay width measurement.

For the $U(1)_{\mathbf{A} Y + \mathbf{B} q}$ model, we compute the $Z$ decay width with two different $g_x$ values we use in the discussion of dark matter in Section.~\ref{sec:DM}.
For each $g_x$ value,
we evaluate two edge points on the plot of Fig.~\ref{fig:WMG} which could achieve 77~MeV $W$ boson mass enhancement.
For $g_x=0.05$,
%take two different sets of parameters, $\{\alpha=0.51,M_1=600\text{GeV}\},\{\alpha=1.7,M_1=2000\text{GeV}\}$,
\begin{align}
    \Gamma_Z(\alpha=0.51,M_1=600~\text{GeV})\approx&\,2.4933~\text{GeV},
    \left|\Gamma_Z^{\mathbf{A} Y + \mathbf{B} q}-\Gamma_Z^{EX}\right|\approx 0.83~\text{errorbar}\\
     \Gamma_Z(\alpha=1.7,M_1=2000~\text{GeV})\approx&\,2.4934~\text{GeV},
     \left|\Gamma_Z^{\mathbf{A} Y + \mathbf{B} q}-\Gamma_Z^{EX}\right|\approx 0.79~\text{errorbar}\, ,
\end{align}
for $g_x=0.08$,
\begin{align}
    \Gamma_Z(\alpha=0.51,M_1=600~\text{GeV})\approx&\,2.4930~\text{GeV}, \left|\Gamma_Z^{\mathbf{A} Y + \mathbf{B} q}-\Gamma_Z^{EX}\right|\approx 0.97~\text{errorbar}\\
    \Gamma_Z(\alpha=1.7,M_1=2000~\text{GeV})\approx&\,2.4930~\text{GeV},\left|\Gamma_Z^{\mathbf{A} Y + \mathbf{B} q}-\Gamma_Z^{EX}\right|\approx 0.93~\text{errorbar}\, ,
\end{align}
where $\Gamma_Z^{EX}\approx 2.4952\text{GeV}\pm 0.0023\text{GeV}$~\cite{ParticleDataGroup:2018ovx}.
Comparing to the $Z$ boson decay width,
other precision electroweak data sets weaker constraints to our models.
A detail analysis was done in, for example~\cite{Feldman:2007wj,Feldman:2006wb},
showing their model fits quite well to the experimental data,
in which a much larger $U(1)$ mixing was considered compared to the two models we discuss in this paper.
Thus both of our models are safe confronting the precision electroweak data.


\begin{thebibliography}{999}

%\cite{CDF:2022hxs}
\bibitem{CDF:2022hxs}
T.~Aaltonen \textit{et al.} [CDF],
%``High-precision measurement of the W boson mass with the CDF II detector,''
Science \textbf{376}, no.6589, 170-176 (2022)
doi:10.1126/science.abk1781
%25 citations counted in INSPIRE as of 12 Apr 2022

%\cite{ParticleDataGroup:2020ssz}
\bibitem{ParticleDataGroup:2020ssz}
P.~A.~Zyla \textit{et al.} [Particle Data Group],
%``Review of Particle Physics,''
PTEP \textbf{2020}, no.8, 083C01 (2020)
doi:10.1093/ptep/ptaa104
%3299 citations counted in INSPIRE as of 12 Apr 2022

%%%%%%%%%%%%%%%%%%%%%%%%%%

%\cite{Fan:2022dck}
\bibitem{Fan:2022dck}
Y.~Z.~Fan, T.~P.~Tang, Y.~L.~S.~Tsai and L.~Wu,
%``Inert Higgs Dark Matter for New CDF W-boson Mass and Detection Prospects,''
[arXiv:2204.03693 [hep-ph]].
%6 citations counted in INSPIRE as of 13 Apr 2022

%\cite{Zhu:2022tpr}
\bibitem{Zhu:2022tpr}
C.~R.~Zhu, M.~Y.~Cui, Z.~Q.~Xia, Z.~H.~Yu, X.~Huang, Q.~Yuan and Y.~Z.~Fan,
%``GeV antiproton/gamma-ray excesses and the $W$-boson mass anomaly: three faces of $\sim 60-70$ GeV dark matter particle?,''
[arXiv:2204.03767 [astro-ph.HE]].
%2 citations counted in INSPIRE as of 13 Apr 2022

%\cite{Lu:2022bgw}
\bibitem{Lu:2022bgw}
C.~T.~Lu, L.~Wu, Y.~Wu and B.~Zhu,
%``Electroweak Precision Fit and New Physics in light of $W$ Boson Mass,''
[arXiv:2204.03796 [hep-ph]].
%14 citations counted in INSPIRE as of 14 Apr 2022

%\cite{Athron:2022qpo}
\bibitem{Athron:2022qpo}
P.~Athron, A.~Fowlie, C.~T.~Lu, L.~Wu, Y.~Wu and B.~Zhu,
%``The $W$ boson Mass and Muon $g-2$: Hadronic Uncertainties or New Physics?,''
[arXiv:2204.03996 [hep-ph]].
%6 citations counted in INSPIRE as of 13 Apr 2022

%\cite{Yuan:2022cpw}
\bibitem{Yuan:2022cpw}
G.~W.~Yuan, L.~Zu, L.~Feng and Y.~F.~Cai,
%``$W$-boson mass anomaly: probing the models of axion-like particle, dark photon and Chameleon dark energy,''
[arXiv:2204.04183 [hep-ph]].
%11 citations counted in INSPIRE as of 14 Apr 2022

%\cite{Strumia:2022qkt}
\bibitem{Strumia:2022qkt}
A.~Strumia,
%``Interpreting electroweak precision data including the $W$-mass CDF anomaly,''
[arXiv:2204.04191 [hep-ph]].
%7 citations counted in INSPIRE as of 13 Apr 2022

%\cite{Yang:2022gvz}
\bibitem{Yang:2022gvz}
J.~M.~Yang and Y.~Zhang,
%``Low energy SUSY confronted with new measurements of W-boson mass and muon g-2,''
[arXiv:2204.04202 [hep-ph]].
%5 citations counted in INSPIRE as of 13 Apr 2022

%\cite{deBlas:2022hdk}
\bibitem{deBlas:2022hdk}
J.~de Blas, M.~Pierini, L.~Reina and L.~Silvestrini,
%``Impact of the recent measurements of the top-quark and W-boson masses on electroweak precision fits,''
[arXiv:2204.04204 [hep-ph]].
%6 citations counted in INSPIRE as of 13 Apr 2022

%\cite{Du:2022pbp}
\bibitem{Du:2022pbp}
X.~K.~Du, Z.~Li, F.~Wang and Y.~K.~Zhang,
%``Explaining The Muon $g-2$ Anomaly and New CDFII W-Boson Mass in the Framework of ExtraOrdinary Gauge Mediation,''
[arXiv:2204.04286 [hep-ph]].
%0 citations counted in INSPIRE as of 13 Apr 2022

%\cite{Tang:2022pxh}
\bibitem{Tang:2022pxh}
T.~P.~Tang, M.~Abdughani, L.~Feng, Y.~L.~S.~Tsai and Y.~Z.~Fan,
%``NMSSM neutralino dark matter for $W$-boson mass and muon $g-2$ and the promising prospect of direct detection,''
[arXiv:2204.04356 [hep-ph]].
%0 citations counted in INSPIRE as of 13 Apr 2022

%\cite{Cacciapaglia:2022xih}
\bibitem{Cacciapaglia:2022xih}
G.~Cacciapaglia and F.~Sannino,
%``The W boson mass weighs in on the non-standard Higgs,''
[arXiv:2204.04514 [hep-ph]].
%0 citations counted in INSPIRE as of 13 Apr 2022

%\cite{Blennow:2022yfm}
\bibitem{Blennow:2022yfm}
M.~Blennow, P.~Coloma, E.~Fern\'andez-Mart\'\i{}nez and M.~Gonz\'alez-L\'opez,
%``Right-handed neutrinos and the CDF II anomaly,''
[arXiv:2204.04559 [hep-ph]].
%0 citations counted in INSPIRE as of 13 Apr 2022

%\cite{Arias-Aragon:2022ats}
\bibitem{Arias-Aragon:2022ats}
F.~Arias-Arag\'on, E.~Fern\'andez-Mart\'\i{}nez, M.~Gonz\'alez-L\'opez and L.~Merlo,
%``Dynamical Minimal Flavour Violating Inverse Seesaw,''
[arXiv:2204.04672 [hep-ph]].
%0 citations counted in INSPIRE as of 13 Apr 2022

%\cite{Zhu:2022scj}
\bibitem{Zhu:2022scj}
B.~Y.~Zhu, S.~Li, J.~G.~Cheng, R.~L.~Li and Y.~F.~Liang,
%``Using gamma-ray observation of dwarf spheroidal galaxy to test a dark matter model that can interpret the W-boson mass anomaly,''
[arXiv:2204.04688 [astro-ph.HE]].
%0 citations counted in INSPIRE as of 13 Apr 2022

%\cite{Sakurai:2022hwh}
\bibitem{Sakurai:2022hwh}
K.~Sakurai, F.~Takahashi and W.~Yin,
%``Singlet extensions and W boson mass in the light of the CDF II result,''
[arXiv:2204.04770 [hep-ph]].
%0 citations counted in INSPIRE as of 13 Apr 2022

%\cite{Fan:2022yly}
\bibitem{Fan:2022yly}
J.~Fan, L.~Li, T.~Liu and K.~F.~Lyu,
%``$W$-Boson Mass, Electroweak Precision Tests and SMEFT,''
[arXiv:2204.04805 [hep-ph]].
%0 citations counted in INSPIRE as of 13 Apr 2022

%\cite{Liu:2022jdq}
\bibitem{Liu:2022jdq}
X.~Liu, S.~Y.~Guo, B.~Zhu and Y.~Li,
%``Unifying gravitational waves with $W$ boson, FIMP dark matter, and Majorana Seesaw mechanism,''
[arXiv:2204.04834 [hep-ph]].
%5 citations counted in INSPIRE as of 14 Apr 2022

%\cite{Lee:2022nqz}
\bibitem{Lee:2022nqz}
H.~M.~Lee and K.~Yamashita,
%``A Model of Vector-like Leptons for the Muon $g-2$ and the $W$ Boson Mass,''
[arXiv:2204.05024 [hep-ph]].
%6 citations counted in INSPIRE as of 14 Apr 2022

%\cite{Cheng:2022jyi}
\bibitem{Cheng:2022jyi}
Y.~Cheng, X.~G.~He, Z.~L.~Huang and M.~W.~Li,
%``Type-II Seesaw Triplet Scalar and Its VEV Effects on Neutrino Trident Scattering and W mass,''
[arXiv:2204.05031 [hep-ph]].
%0 citations counted in INSPIRE as of 13 Apr 2022

%\cite{Song:2022xts}
\bibitem{Song:2022xts}
H.~Song, W.~Su and M.~Zhang,
%``Electroweak Phase Transition in 2HDM under Higgs, Z-pole, and W precision measurements,''
[arXiv:2204.05085 [hep-ph]].
%0 citations counted in INSPIRE as of 13 Apr 2022

%\cite{Bagnaschi:2022whn}
\bibitem{Bagnaschi:2022whn}
E.~Bagnaschi, J.~Ellis, M.~Madigan, K.~Mimasu, V.~Sanz and T.~You,
%``SMEFT Analysis of $m_{W}$,''
[arXiv:2204.05260 [hep-ph]].
%0 citations counted in INSPIRE as of 13 Apr 2022

%\cite{Paul:2022dds}
\bibitem{Paul:2022dds}
A.~Paul and M.~Valli,
%``Violation of custodial symmetry from W-boson mass measurements,''
[arXiv:2204.05267 [hep-ph]].
%0 citations counted in INSPIRE as of 13 Apr 2022

%\cite{Bahl:2022xzi}
\bibitem{Bahl:2022xzi}
H.~Bahl, J.~Braathen and G.~Weiglein,
%``New physics effects on the $W$-boson mass from a doublet extension of the SM Higgs sector,''
[arXiv:2204.05269 [hep-ph]].
%0 citations counted in INSPIRE as of 13 Apr 2022

%\cite{Asadi:2022xiy}
\bibitem{Asadi:2022xiy}
P.~Asadi, C.~Cesarotti, K.~Fraser, S.~Homiller and A.~Parikh,
%``Oblique Lessons from the $W$ Mass Measurement at CDF II,''
[arXiv:2204.05283 [hep-ph]].
%0 citations counted in INSPIRE as of 13 Apr 2022

%\cite{DiLuzio:2022xns}
\bibitem{DiLuzio:2022xns}
L.~Di Luzio, R.~Gr\"ober and P.~Paradisi,
%``Higgs physics confronts the $M_W$ anomaly,''
[arXiv:2204.05284 [hep-ph]].
%0 citations counted in INSPIRE as of 13 Apr 2022

%\cite{Athron:2022isz}
\bibitem{Athron:2022isz}
P.~Athron, M.~Bach, D.~H.~J.~Jacob, W.~Kotlarski, D.~St\"ockinger and A.~Voigt,
%``Precise calculation of the W boson pole mass beyond the Standard Model with FlexibleSUSY,''
[arXiv:2204.05285 [hep-ph]].
%0 citations counted in INSPIRE as of 13 Apr 2022

%\cite{Gu:2022htv}
\bibitem{Gu:2022htv}
J.~Gu, Z.~Liu, T.~Ma and J.~Shu,
%``Speculations on the W-Mass Measurement at CDF,''
[arXiv:2204.05296 [hep-ph]].
%4 citations counted in INSPIRE as of 14 Apr 2022

%\cite{Heckman:2022the}
\bibitem{Heckman:2022the}
J.~J.~Heckman,
%``Extra $W$-Boson Mass from a D3-Brane,''
[arXiv:2204.05302 [hep-ph]].
%0 citations counted in INSPIRE as of 13 Apr 2022

%\cite{Babu:2022pdn}
\bibitem{Babu:2022pdn}
K.~S.~Babu, S.~Jana and V.~P.~K.,
%``Correlating $W$-Boson Mass Shift with Muon \textbackslash{}boldmath${g-2}$ in the 2HDM,''
[arXiv:2204.05303 [hep-ph]].
%0 citations counted in INSPIRE as of 13 Apr 2022

%\cite{Heo:2022dey}
\bibitem{Heo:2022dey}
Y.~Heo, D.~W.~Jung and J.~S.~Lee,
%``Impact of the CDF $W$-mass anomaly on two Higgs doublet model,''
[arXiv:2204.05728 [hep-ph]].
%2 citations counted in INSPIRE as of 14 Apr 2022

%\cite{Du:2022brr}
\bibitem{Du:2022brr}
X.~K.~Du, Z.~Li, F.~Wang and Y.~K.~Zhang,
%``Explaining The New CDFII W-Boson Mass In The Georgi-Machacek Extension Models,''
[arXiv:2204.05760 [hep-ph]].
%1 citations counted in INSPIRE as of 14 Apr 2022

%\cite{Cheung:2022zsb}
\bibitem{Cheung:2022zsb}
K.~Cheung, W.~Y.~Keung and P.~Y.~Tseng,
%``Iso-doublet Vector Leptoquark solution to the Muon $g-2$, $R_{K, K^*}$, $R_{D,D^*}$, and $W$-mass Anomalies,''
[arXiv:2204.05942 [hep-ph]].
%1 citations counted in INSPIRE as of 14 Apr 2022

%\cite{DiLuzio:2022ziu}
\bibitem{DiLuzio:2022ziu}
L.~Di Luzio, M.~Nardecchia and C.~Toni,
%``Light vectors coupled to anomalous currents with harmless Wess-Zumino terms,''
[arXiv:2204.05945 [hep-ph]].
%0 citations counted in INSPIRE as of 14 Apr 2022

%\cite{Crivellin:2022fdf}
\bibitem{Crivellin:2022fdf}
A.~Crivellin, M.~Kirk, T.~Kitahara and F.~Mescia,
%``Correlating $t\to cZ$ to the $W$ Mass and $B$ Physics with Vector-Like Quarks,''
[arXiv:2204.05962 [hep-ph]].
%1 citations counted in INSPIRE as of 14 Apr 2022

%\cite{Endo:2022kiw}
\bibitem{Endo:2022kiw}
M.~Endo and S.~Mishima,
%``New physics interpretation of $W$-boson mass anomaly,''
[arXiv:2204.05965 [hep-ph]].
%1 citations counted in INSPIRE as of 14 Apr 2022

%\cite{Biekotter:2022abc}
\bibitem{Biekotter:2022abc}
T.~Biek\"otter, S.~Heinemeyer and G.~Weiglein,
%``Excesses in the low-mass Higgs-boson search and the W-boson mass measurement,''
[arXiv:2204.05975 [hep-ph]].
%1 citations counted in INSPIRE as of 14 Apr 2022

%\cite{Balkin:2022glu}
\bibitem{Balkin:2022glu}
R.~Balkin, E.~Madge, T.~Menzo, G.~Perez, Y.~Soreq and J.~Zupan,
%``On the implications of positive W mass shift,''
[arXiv:2204.05992 [hep-ph]].
%1 citations counted in INSPIRE as of 14 Apr 2022

%\cite{Krasnikov:2022xsi}
\bibitem{Krasnikov:2022xsi}
N.~V.~Krasnikov,
%``Nonlocal generalization of the SM as an explanation of recent CDF result,''
[arXiv:2204.06327 [hep-ph]].
%0 citations counted in INSPIRE as of 14 Apr 2022

%\cite{Ahn:2022xeq}
\bibitem{Ahn:2022xeq}
Y.~H.~Ahn, S.~K.~Kang and R.~Ramos,
%``Implications of New CDF-II $W$ Boson Mass on Two Higgs Doublet Model,''
[arXiv:2204.06485 [hep-ph]].
%0 citations counted in INSPIRE as of 14 Apr 2022

%\cite{Han:2022juu}
\bibitem{Han:2022juu}
X.~F.~Han, F.~Wang, L.~Wang, J.~M.~Yang and Y.~Zhang,
%``A joint explanation of W-mass and muon g-2 in 2HDM,''
[arXiv:2204.06505 [hep-ph]].
%0 citations counted in INSPIRE as of 14 Apr 2022

%\cite{Zheng:2022irz}
\bibitem{Zheng:2022irz}
M.~D.~Zheng, F.~Z.~Chen and H.~H.~Zhang,
%``The $W\ell\nu$-vertex corrections to W-boson mass in the R-parity violating MSSM,''
[arXiv:2204.06541 [hep-ph]].
%0 citations counted in INSPIRE as of 14 Apr 2022

%\cite{Kawamura:2022uft}
\bibitem{Kawamura:2022uft}
J.~Kawamura, S.~Okawa and Y.~Omura,
%``$W$ boson mass and muon $g-2$ in a lepton portal dark matter model,''
[arXiv:2204.07022 [hep-ph]].
%0 citations counted in INSPIRE as of 15 Apr 2022

%\cite{Peli:2022ybi}
\bibitem{Peli:2022ybi}
Z.~P\'eli and Z.~Tr\'ocs\'anyi,
%``Vacuum stability and scalar masses in the superweak extension of the standard model,''
[arXiv:2204.07100 [hep-ph]].
%0 citations counted in INSPIRE as of 15 Apr 2022

%\cite{Ghoshal:2022vzo}
\bibitem{Ghoshal:2022vzo}
A.~Ghoshal, N.~Okada, S.~Okada, D.~Raut, Q.~Shafi and A.~Thapa,
%``Type III seesaw with R-parity violation in light of $m_W$ (CDF),''
[arXiv:2204.07138 [hep-ph]].
%0 citations counted in INSPIRE as of 15 Apr 2022

%\cite{Perez:2022uil}
\bibitem{Perez:2022uil}
P.~F.~Perez, H.~H.~Patel and A.~D.~Plascencia,
%``On the $W$-mass and New Higgs Bosons,''
[arXiv:2204.07144 [hep-ph]].
%0 citations counted in INSPIRE as of 15 Apr 2022

%\cite{Kanemura:2022ahw}
\bibitem{Kanemura:2022ahw}
S.~Kanemura and K.~Yagyu,
%``Implication of the $W$ boson mass anomaly at CDF II in the Higgs triplet model with a mass difference,''
[arXiv:2204.07511 [hep-ph]].
%6 citations counted in INSPIRE as of 20 Apr 2022

%%%%%%%%%%%%%%%%%%%%%%%%%%%%%%%%%%%

%\cite{D0:2012kms}
\bibitem{D0:2012kms}
V.~M.~Abazov \textit{et al.} [D0],
%``Measurement of the W Boson Mass with the D0 Detector,''
Phys. Rev. Lett. \textbf{108}, 151804 (2012)
doi:10.1103/PhysRevLett.108.151804
[arXiv:1203.0293 [hep-ex]].
%148 citations counted in INSPIRE as of 13 Apr 2022

%\cite{ALEPH:2013dgf}
\bibitem{ALEPH:2013dgf}
S.~Schael \textit{et al.} [ALEPH, DELPHI, L3, OPAL and LEP Electroweak],
%``Electroweak Measurements in Electron-Positron Collisions at W-Boson-Pair Energies at LEP,''
Phys. Rept. \textbf{532}, 119-244 (2013)
doi:10.1016/j.physrep.2013.07.004
[arXiv:1302.3415 [hep-ex]].
%633 citations counted in INSPIRE as of 13 Apr 2022

%\cite{ATLAS:2017rzl}
\bibitem{ATLAS:2017rzl}
M.~Aaboud \textit{et al.} [ATLAS],
%``Measurement of the $W$-boson mass in pp collisions at $\sqrt{s}=7$ TeV with the ATLAS detector,''
Eur. Phys. J. C \textbf{78}, no.2, 110 (2018)
[erratum: Eur. Phys. J. C \textbf{78}, no.11, 898 (2018)]
doi:10.1140/epjc/s10052-017-5475-4
[arXiv:1701.07240 [hep-ex]].
%306 citations counted in INSPIRE as of 13 Apr 2022

%\cite{LHCb:2021bjt}
\bibitem{LHCb:2021bjt}
R.~Aaij \textit{et al.} [LHCb],
%``Measurement of the W boson mass,''
JHEP \textbf{01}, 036 (2022)
doi:10.1007/JHEP01(2022)036
[arXiv:2109.01113 [hep-ex]].
%17 citations counted in INSPIRE as of 13 Apr 2022

%\cite{Feng:2014cla}
\bibitem{Feng:2014cla}
W.~Z.~Feng, G.~Shiu, P.~Soler and F.~Ye,
%``Building a St\"uckelberg portal,''
JHEP \textbf{05}, 065 (2014)
doi:10.1007/JHEP05(2014)065
[arXiv:1401.5890 [hep-ph]].
%36 citations counted in INSPIRE as of 20 Apr 2022

%\cite{Feng:2014eja}
\bibitem{Feng:2014eja}
W.~Z.~Feng, G.~Shiu, P.~Soler and F.~Ye,
%``Probing Hidden Sectors with St\"uckelberg U(1) Gauge Fields,''
Phys. Rev. Lett. \textbf{113}, 061802 (2014)
doi:10.1103/PhysRevLett.113.061802
[arXiv:1401.5880 [hep-ph]].
%31 citations counted in INSPIRE as of 20 Apr 2022

%\cite{Anchordoqui:2014wha}
\bibitem{Anchordoqui:2014wha}
L.~A.~Anchordoqui, I.~Antoniadis, D.~C.~Dai, W.~Z.~Feng, H.~Goldberg, X.~Huang, D.~Lust, D.~Stojkovic and T.~R.~Taylor,
%``String Resonances at Hadron Colliders,''
Phys. Rev. D \textbf{90}, no.6, 066013 (2014)
doi:10.1103/PhysRevD.90.066013
[arXiv:1407.8120 [hep-ph]].
%32 citations counted in INSPIRE as of 20 Apr 2022

%\cite{Holdom:1985ag}
\bibitem{Holdom:1985ag}
B.~Holdom,
%``Two U(1)'s and Epsilon Charge Shifts,''
Phys. Lett. B \textbf{166}, 196-198 (1986)
doi:10.1016/0370-2693(86)91377-8
%2015 citations counted in INSPIRE as of 15 Apr 2022

%\cite{Aboubrahim:2020lnr}
\bibitem{Aboubrahim:2020lnr}
A.~Aboubrahim, W.~Z.~Feng, P.~Nath and Z.~Y.~Wang,
%``Self-interacting hidden sector dark matter, small scale galaxy structure anomalies, and a dark force,''
Phys. Rev. D \textbf{103}, no.7, 075014 (2021)
doi:10.1103/PhysRevD.103.075014
[arXiv:2008.00529 [hep-ph]].
%9 citations counted in INSPIRE as of 15 Apr 2022

%\cite{Feldman:2007wj}
\bibitem{Feldman:2007wj}
D.~Feldman, Z.~Liu and P.~Nath,
%``The Stueckelberg Z-prime Extension with Kinetic Mixing and Milli-Charged Dark Matter From the Hidden Sector,''
Phys. Rev. D \textbf{75}, 115001 (2007)
doi:10.1103/PhysRevD.75.115001
[arXiv:hep-ph/0702123 [hep-ph]].
%325 citations counted in INSPIRE as of 15 Apr 2022

%\cite{Kors:2004dx}
\bibitem{Kors:2004dx}
B.~Kors and P.~Nath,
%``A Stueckelberg extension of the standard model,''
Phys. Lett. B \textbf{586}, 366-372 (2004)
doi:10.1016/j.physletb.2004.02.051
[arXiv:hep-ph/0402047 [hep-ph]].
%215 citations counted in INSPIRE as of 15 Apr 2022

%\cite{Kors:2005uz}
\bibitem{Kors:2005uz}
B.~Kors and P.~Nath,
%``Aspects of the Stueckelberg extension,''
JHEP \textbf{07}, 069 (2005)
doi:10.1088/1126-6708/2005/07/069
[arXiv:hep-ph/0503208 [hep-ph]].
%165 citations counted in INSPIRE as of 15 Apr 2022

%\cite{Feldman:2006wb}
\bibitem{Feldman:2006wb}
D.~Feldman, Z.~Liu and P.~Nath,
%``The Stueckelberg $Z$ Prime at the LHC: Discovery Potential, Signature Spaces and Model Discrimination,''
JHEP \textbf{11}, 007 (2006)
doi:10.1088/1126-6708/2006/11/007
[arXiv:hep-ph/0606294 [hep-ph]].
%106 citations counted in INSPIRE as of 15 Apr 2022

%\cite{Feng:2012jn}
\bibitem{Feng:2012jn}
W.~Z.~Feng, P.~Nath and G.~Peim,
%``Cosmic Coincidence and Asymmetric Dark Matter in a Stueckelberg Extension,''
Phys. Rev. D \textbf{85}, 115016 (2012)
doi:10.1103/PhysRevD.85.115016
[arXiv:1204.5752 [hep-ph]].
%47 citations counted in INSPIRE as of 17 Apr 2022

%\cite{Celis:2016ayl}
\bibitem{Celis:2016ayl}
A.~Celis, W.~Z.~Feng and M.~Vollmann,
%``Dirac dark matter and $b \to s \ell^+ \ell^-$ with $\mathrm{U(1)}$ gauge symmetry,''
Phys. Rev. D \textbf{95}, no.3, 035018 (2017)
doi:10.1103/PhysRevD.95.035018
[arXiv:1608.03894 [hep-ph]].
%43 citations counted in INSPIRE as of 26 Apr 2022

%\cite{Peskin:1990zt}
\bibitem{Peskin:1990zt}
M.~E.~Peskin and T.~Takeuchi,
%``A New constraint on a strongly interacting Higgs sector,''
Phys. Rev. Lett. \textbf{65}, 964-967 (1990)
doi:10.1103/PhysRevLett.65.964
%2035 citations counted in INSPIRE as of 13 Apr 2022

%\cite{Peskin:1991sw}
\bibitem{Peskin:1991sw}
M.~E.~Peskin and T.~Takeuchi,
%``Estimation of oblique electroweak corrections,''
Phys. Rev. D \textbf{46}, 381-409 (1992)
doi:10.1103/PhysRevD.46.381
%2422 citations counted in INSPIRE as of 13 Apr 2022

%\cite{Holdom:1990xp}
\bibitem{Holdom:1990xp}
B.~Holdom,
%``Oblique electroweak corrections and an extra gauge boson,''
Phys. Lett. B \textbf{259}, 329-334 (1991)
doi:10.1016/0370-2693(91)90836-F
%189 citations counted in INSPIRE as of 15 Apr 2022

%\cite{Burgess:1993vc}
\bibitem{Burgess:1993vc}
C.~P.~Burgess, S.~Godfrey, H.~Konig, D.~London and I.~Maksymyk,
%``Model independent global constraints on new physics,''
Phys. Rev. D \textbf{49}, 6115-6147 (1994)
doi:10.1103/PhysRevD.49.6115
[arXiv:hep-ph/9312291 [hep-ph]].
%201 citations counted in INSPIRE as of 13 Apr 2022

%\cite{Babu:1997st}
\bibitem{Babu:1997st}
K.~S.~Babu, C.~F.~Kolda and J.~March-Russell,
%``Implications of generalized Z - Z-prime mixing,''
Phys. Rev. D \textbf{57}, 6788-6792 (1998)
doi:10.1103/PhysRevD.57.6788
[arXiv:hep-ph/9710441 [hep-ph]].
%316 citations counted in INSPIRE as of 13 Apr 2022




%\cite{ATLAS:2019erb}
\bibitem{ATLAS:2019erb}
G.~Aad \textit{et al.} [ATLAS],
%``Search for high-mass dilepton resonances using 139 fb$^{-1}$ of $pp$ collision data collected at $\sqrt{s}=$13 TeV with the ATLAS detector,''
Phys. Lett. B \textbf{796}, 68-87 (2019)
doi:10.1016/j.physletb.2019.07.016
[arXiv:1903.06248 [hep-ex]].
%253 citations counted in INSPIRE as of 12 Apr 2022

%\cite{Fairbairn:2016iuf}
\bibitem{Fairbairn:2016iuf}
M.~Fairbairn, J.~Heal, F.~Kahlhoefer and P.~Tunney,
%``Constraints on Z' models from LHC dijet searches and implications for dark matter,''
JHEP \textbf{09}, 018 (2016)
doi:10.1007/JHEP09(2016)018
[arXiv:1605.07940 [hep-ph]].
%57 citations counted in INSPIRE as of 14 Nov 2022

%\cite{ParticleDataGroup:2018ovx}
\bibitem{ParticleDataGroup:2018ovx}
M.~Tanabashi \textit{et al.} [Particle Data Group],
%``Review of Particle Physics,''
Phys. Rev. D \textbf{98}, no.3, 030001 (2018)
doi:10.1103/PhysRevD.98.030001
%7691 citations counted in INSPIRE as of 12 Apr 2022

%\cite{Muong-2:2021ojo}
\bibitem{Muong-2:2021ojo}
B.~Abi \textit{et al.} [Muon g-2],
%``Measurement of the Positive Muon Anomalous Magnetic Moment to 0.46 ppm,''
Phys. Rev. Lett. \textbf{126}, no.14, 141801 (2021)
doi:10.1103/PhysRevLett.126.141801
[arXiv:2104.03281 [hep-ex]].
%696 citations counted in INSPIRE as of 12 Apr 2022

%\cite{Muong-2:2006rrc}
\bibitem{Muong-2:2006rrc}
G.~W.~Bennett \textit{et al.} [Muon g-2],
%``Final Report of the Muon E821 Anomalous Magnetic Moment Measurement at BNL,''
Phys. Rev. D \textbf{73}, 072003 (2006)
doi:10.1103/PhysRevD.73.072003
[arXiv:hep-ex/0602035 [hep-ex]].
%2824 citations counted in INSPIRE as of 13 Apr 2022

%\cite{Aoyama:2020ynm}
\bibitem{Aoyama:2020ynm}
T.~Aoyama, N.~Asmussen, M.~Benayoun, J.~Bijnens, T.~Blum, M.~Bruno, I.~Caprini, C.~M.~Carloni Calame, M.~C\`e and G.~Colangelo, \textit{et al.}
%``The anomalous magnetic moment of the muon in the Standard Model,''
Phys. Rept. \textbf{887}, 1-166 (2020)
doi:10.1016/j.physrep.2020.07.006
[arXiv:2006.04822 [hep-ph]].
%598 citations counted in INSPIRE as of 12 Apr 2022



%\cite{LUX:2016ggv}
\bibitem{LUX:2016ggv}
D.~S.~Akerib \textit{et al.} [LUX],
%``Results from a search for dark matter in the complete LUX exposure,''
Phys. Rev. Lett. \textbf{118}, no.2, 021303 (2017)
doi:10.1103/PhysRevLett.118.021303
[arXiv:1608.07648 [astro-ph.CO]].
%1665 citations counted in INSPIRE as of 17 Apr 2022

%\cite{XENON:2017vdw}
\bibitem{XENON:2017vdw}
E.~Aprile \textit{et al.} [XENON],
%``First Dark Matter Search Results from the XENON1T Experiment,''
Phys. Rev. Lett. \textbf{119}, no.18, 181301 (2017)
doi:10.1103/PhysRevLett.119.181301
[arXiv:1705.06655 [astro-ph.CO]].
%993 citations counted in INSPIRE as of 17 Apr 2022

%\cite{PandaX-II:2017hlx}
\bibitem{PandaX-II:2017hlx}
X.~Cui \textit{et al.} [PandaX-II],
%``Dark Matter Results From 54-Ton-Day Exposure of PandaX-II Experiment,''
Phys. Rev. Lett. \textbf{119}, no.18, 181302 (2017)
doi:10.1103/PhysRevLett.119.181302
[arXiv:1708.06917 [astro-ph.CO]].
%919 citations counted in INSPIRE as of 17 Apr 2022

%\cite{Planck:2015fie}
\bibitem{Planck:2015fie}
P.~A.~R.~Ade \textit{et al.} [Planck],
%``Planck 2015 results. XIII. Cosmological parameters,''
Astron. Astrophys. \textbf{594}, A13 (2016)
doi:10.1051/0004-6361/201525830
[arXiv:1502.01589 [astro-ph.CO]].
%11172 citations counted in INSPIRE as of 14 Nov 2022

%\cite{AMS:2014xys}
\bibitem{AMS:2014xys}
M.~Aguilar \textit{et al.} [AMS],
%``Electron and Positron Fluxes in Primary Cosmic Rays Measured with the Alpha Magnetic Spectrometer on the International Space Station,''
Phys. Rev. Lett. \textbf{113}, 121102 (2014)
doi:10.1103/PhysRevLett.113.121102
%579 citations counted in INSPIRE as of 14 Nov 2022

%\cite{AMS:2014bun}
\bibitem{AMS:2014bun} L.~Accardo \textit{et al.} [AMS],
%``High Statistics Measurement of the Positron Fraction in Primary Cosmic Rays of 0.5\textendash{}500 GeV with the Alpha Magnetic Spectrometer on the International Space Station,''
Phys. Rev. Lett. \textbf{113}, 121101 (2014)
doi:10.1103/PhysRevLett.113.121101
%631 citations counted in INSPIRE as of 14 Nov 2022

%\cite{Fermi-LAT:2015att}
\bibitem{Fermi-LAT:2015att} M.~Ackermann \textit{et al.} [Fermi-LAT],
%``Searching for Dark Matter Annihilation from Milky Way Dwarf Spheroidal Galaxies with Six Years of Fermi Large Area Telescope Data,''
Phys. Rev. Lett. \textbf{115}, no.23, 231301 (2015)
doi:10.1103/PhysRevLett.115.231301 [arXiv:1503.02641 [astro-ph.HE]].
%1190 citations counted in INSPIRE as of 14 Nov 2022

%\cite{Baur:2001ze}
\bibitem{Baur:2001ze}
U.~Baur, O.~Brein, W.~Hollik, C.~Schappacher and D.~Wackeroth,
%``Electroweak radiative corrections to neutral current Drell-Yan processes at hadron colliders,''
Phys. Rev. D \textbf{65}, 033007 (2002)
doi:10.1103/PhysRevD.65.033007
[arXiv:hep-ph/0108274 [hep-ph]].
%290 citations counted in INSPIRE as of 13 Nov 2022



\end{thebibliography}
\end{document}